\documentclass[11pt]{article}

\usepackage{amsfonts}
\usepackage{epsfig}
\usepackage{latexsym}
\usepackage{amsmath}
\usepackage{amssymb}
\usepackage{float}
\usepackage{graphicx}

\def\bequ{\begin{equation}}
\def\eequ{\end{equation}}
\def\barr{\begin{array}}
\def\earr{\end{array}}

\def\ben{\begin{equation}}
\def\een{\end{equation}}
\def\bena{\begin{eqnarray}}
\def\eena{\end{eqnarray}}

\def\spa#1{\phantom{\fbox{\rule[-#1cm]{0cm}{0cm}}}}

\def\b1{e^0}

\newcommand{\be}{\begin{equation}}
\newcommand{\ee}{\end{equation}}
\def\bea{\begin{eqnarray}}
\def\eea{\end{eqnarray}}

\def\be{\begin{equation}}
\def\ee{\end{equation}}
\def\bea{\begin{eqnarray}}
\def\eea{\end{eqnarray}}

\def\lesssim{\mathrel{\hbox{\rlap{\hbox{\lower4pt\hbox{$\sim$}}}\hbox{$<$}}}}
\def\gtrsim{\mathrel{\hbox{\rlap{\hbox{\lower4pt\hbox{$\sim$}}}\hbox{$>$}}}}

\begin{document}
\title{{\huge \bf{Scalar Casimir Effect on a \\ D-dimensional Einstein Static Universe}}}
\author{{Carlos A. R. Herdeiro$^{1}$\footnote{crherdei@fc.up.pt}, \ Raquel H. Ribeiro$^{1}$\footnote{irhr2@cam.ac.uk}} \ and Marco Sampaio$^{2}$\footnote{sampaio@hep.phy.cam.ac.uk}
\\
\\ {$^{1}${\em Departamento de F\'\i sica e Centro de F\'\i sica do Porto}}
\\ {\em Faculdade de Ci\^encias da
Universidade do Porto}
\\ {\em Rua do Campo Alegre, 687,  4169-007 Porto, Portugal} \spa{0.4cm}\\
{$^{2}${\em Cavendish Laboratory}},
\\ {\em JJ Thomson Avenue},
\\ {\em Cambridge CB3 OHE, United Kingdom}}

\date{November 2007}       
 \maketitle

\begin{abstract}
We compute the renormalised energy momentum tensor of a free scalar field
coupled to gravity  on an
$n+1$-dimensional Einstein Static Universe (ESU), $\mathbb{R}\times
S^n$, with arbitrary low energy effective operators (up to mass
dimension $n+1$). A generic class
of regulators is used, together with the Abel-Plana formula, leading
to a manifestly regulator independent result. The
general structure of the divergences is analysed to show that all the gravitational couplings (not
just the cosmological constant) are renormalised for an arbitrary regulator. Various commonly used methods (damping function, point-splitting, momentum cut-off and zeta function) are shown to, effectively, belong to the given class. The final results depend
strongly on the parity of $n$. A detailed analytical and numerical analysis is performed for the behaviours of the renormalised energy density and a quantity $\sigma$ which determines if the strong energy condition holds for the `quantum fluid'. We briefly discuss the quantum fluid back-reaction problem,  via the higher dimensional Friedmann and Raychaudhuri equations, observe that equilibrium radii exist and unveil the possibility of a `Casimir stabilisation of Einstein Static Universes'. 
\end{abstract}

\newpage

\tableofcontents

\section{Introduction}
The Casimir effect \cite{Casimir} has been under intense experimental
scrutiny over the last ten years, since the landmark experiment of
Lamoreaux \cite{Lamoreaux:1996wh}, and has been experimentally well
established in different setups. A very interesting property of the effect is that it can be either attractive or repulsive, depending sensitively on the boundary conditions, geometry and topology of the background under study \cite{review}.  It would be desirable to have some physical understanding on what, exactly, are the conditions for the effect to be one or the other. In particular because a repulsive Casimir effect may have many interesting applications, as for example, to quantum levitation \cite{ulf}. However such understanding does not exist at present time. 

The issue of a repulsive Casimir effect is particularly timely in the
 context of gravitational physics, since repulsive gravity seems to be dominating our universe at present times \cite{Riess:1998cb,Perlmutter:1998np}. Moreover, repulsive gravity is required to power an early inflationary epoch (see
   \cite{zeldovich} for an example of Casimir induced inflation) and might play a role in avoiding the Big Bang
   singularity, through a bounce (see \cite{Herdeiro:2005zj} for an
   example of a Casimir induced bounce). Thus, any quantum effects that might produce repulsive gravity are of interest, and the Casimir effect is an experimentally confirmed one. The effect has also great relevance in the context of Kaluza-Klein
theories, where the compactness of the extra dimensions confines any
quantum field, and hence produces a Casimir force that tends to make
the extra dimensions perturbatively unstable (as was first argued by
Appelquist and Chodos \cite{Appelquist:1983vs}). In general it seems very hard to find Kaluza-Klein type configurations with compact dimensions which are stable under quantum effects (see \cite{milton}, section 10), without invoking supersymmetry. By the same principle, if our universe has a compact topology, as in a $k=+1$ FRW cosmological model or in $k=0,-1$ cosmological models with identifications, the Casimir effect will induce a force for any quantum field.

Given the lack of general understanding concerning the sign of the Casimir
 force, one might consider families of
 models and examine the change in character of the Casimir
 force as one varies continuous or discrete parameters. Within this
 spirit we study the Casimir effect of a scalar field, $\Phi$,  with
 arbitrary mass $\mu$ and arbitrary coupling $\xi$ to the Ricci scalar of the background on an $n+1$-dimensional Einstein Static Universe (ESU) $\mathbb{R}\times S^n$ of radius $R$. Thus, the model under study is a 4-parameter family, labelled by one discrete, $n$, and three continuous parameters, $\mu,\xi,R$.\footnote{Other studies of the Casimir effect on $n$-dimensional spheres include \cite{Bender:1994zr} (see also \cite{milton}, chapter 9) for a scalar field with $n$-dimensional spherical Dirichlet boundary conditions and \cite{milton} (chapter 10 and references therein) for a scalar field in a Kaluza-Klein setup of type $M^4\times S^n$.}  We will present closed 
 expressions for the renormalised energy density $\rho^{(n)}_{ren}$ and pressure $p^{(n)}_{ren}$, as functions of the aforementioned parameters. These are obtained through a generic regularisation procedure, which can, effectively, be connected to several well known methods such as  point splitting, ``momentum'' cut-off, zeta function and different types of damping functions, including the one discussed in \cite{Ford2}. The mathematical manipulations are done by using the Abel-Plana formula.

Let us note that special cases of our analysis have been performed in the literature. For instance, ref. \cite{Dowker:1976pr} considers the $n=3$ case and conformal coupling; ref. \cite{ozcan} considers general $n$, but only conformal coupling and zero mass. In both cases the results in these references agree with ours. In the first case, a self consistent Einstein Static Universe is found, taking into account the quantum effect; we find similar solutions (see section 4). In the second case, our computation reproduces the result of \cite{ozcan} in the appropriate limit (see (\ref{rhooddconf}) and discussion thereafter).

Besides the intrinsic interest in understanding the behaviour of the Casimir force on a $D$-dimensional Einstein Static Universe we will be also interested in a particular physical application. It is well known that ESUs are, classically, perturbatively unstable towards radial perturbations. One may then ask if the extra Casimir force induced by the quantum field could give rise to stable equilibrium radii. To tackle this question one should use the semi-classical Einstein equations
\[
G_{\mu\nu}=\kappa^2\left(T_{\mu\nu}^{cla}+\langle T_{\mu\nu}^{\Phi}\rangle\right) \ , \]
which yield the higher dimensional Friedmann and Raychaudhuri equations (see e.g. \cite{Garcia:2007gt})
\bequ
\frac{\dot{R}^2+1}{R^2}=\frac{2\kappa^2}{n(n-1)}\left[\rho^{(n)}_{cla}+\rho^{(n)}_{ren}\right] \ , \label{cosmo1}\eequ
with the subscript `cla' denoting the classical quantity, and
\bequ
\frac{\ddot{R}(t)}{R(t)}=-\kappa^2\frac{(n-2)}{n(n-1)}\left[\sigma^{(n)}_{cla}+\sigma^{(n)}_{ren}\right] \ , \label{cosmo2}\eequ
where $\kappa^2=8\pi G_N^{(n+1)}$ and 
\bequ
\sigma^{(n)}\equiv \rho^{(n)}+\frac{n}{n-2}p^{(n)}\ . \label{sigma} \eequ
Thus, in this paper, we will focus our analysis on the renormalised quantities
$\rho^{(n)}_{ren}$ and $\sigma^{(n)}_{ren}$;  the latter determines if
the Casimir effect gives an attractive or repulsive contribution to
the gravitational force. In a follow up paper \cite{paper2} we will examine in detail the dynamics of \eqref{cosmo1} and \eqref{cosmo2}.

This paper is organised as follows. In section 2 we present the theory under study, the unrenormalised quantities and the renormalisation technique. We are then led to equations \eqref{rho_ren1}-\eqref{p_ren2}, which  are the main technical results in this paper. To gain insight about these very generic expressions we analyse their behaviour in section 3, by presenting numerical plots. In section 4 we present a summary of our results and observe that equilibrium radii exist, even including the quantum contribution; we also suggest the possible existence  of \textit{stable} equilibrium radii based on the results herein and \eqref{cosmo1} and \eqref{cosmo2}, an issue to be studied more thoroughly in \cite{paper2}.  In the appendices we discuss some technicalities used in this paper and show how the general regularisation scheme used makes contact with commonly used regularisation methods (point-splitting, ``momentum'' cut-off and zeta function).

\section{Renormalised $T_{\mu \nu}$ for a Scalar Field on ESUs}\label{section_ren_T}
\subsection{General considerations}
We will deal with a simplified model containing the gravitational
field, $g_{\mu\nu}$, and a real scalar field, $\Phi$. We assume that we are at
sufficiently low energy so that these are the only dynamically relevant entities which emerge from a full theory, by integration of any other (more massive)
degrees of freedom. Thus, the low energy effective
action considered herein contains all possible combinations of these fields as
allowed by the surviving symmetries. The coefficient
of each term is unknown, a priory, and arises from the
full theory. However we can still make some considerations about
magnitudes by identifying the relevant scales in the full
theory and in the problem at hand.

The two scales to be considered are the Planck length (which
is assumed to be the fundamental scale in the full theory) and the
largest length scale (or lowest energy scale) in the limit we are
studying. All other dimensionful parameters in the problem should be in
between. We denote the former by $L_{Pl}$ and the latter
by $R$. The most general Lorentz invariant effective action containing $g_{\mu \nu}$ and $\Phi$ in ($n$+1) dimensions will be given by, 
\begin{equation}\label{general_action}
S_{\Phi,g_{\mu\nu}}=\int d^{n+1}\tilde{x}\sum_{j}{\alpha_{j}\left(\frac{L_{j}}{R}\right)^{p_j+q_j}\tilde{\mathcal{R}}^{(p_j)}_{\mu_1,...,\mu_{k_j}}\mathcal{O}^{\mu_1,...,\mu_{k_j}}_{(q_j)}\left(\tilde{D},\tilde{\Phi}\right)} \ , 
\end{equation}
where all lengths are measured in units of $R$ (which is why they are
denoted with a tilde) and the fields have
also been redefined in units of the appropriate quantity constructed
from $R$. $\mathcal{O}^{\ldots}_{(q_j)}$ denotes an arbitrary tensor operator constructed from
$\Phi$ and its derivatives, with (mass) dimension $q_j$,  and $\mathcal{R}^{(p_j)}_{\ldots}$ is an arbitrary tensor constructed from the metric and its derivatives, with dimension $p_j$. The sum over $j$ represents all possibilities for $\{p_j,q_j,k_j\}$, i.e all possible Lorentz invariant combinations of $\mathcal{R}^{(p_j)}_{\ldots}$'s and $\mathcal{O}^{\ldots}_{(q_j)}$'s.  Finally $L_{j}$ is a length scale for
the order $j$ operator, which must be restricted by $L_{Pl}\le
L_j\le R$ and  $\alpha_j$ is a dimensionless coefficient.

 If all scales in the fundamental theory are of the same order as the
Planck scale, then the $L_{j}'s$ are of order $L_{Pl}$. However, more realistically we should expect other larger scales at smaller energies. These may be generated by the
dynamics of the theory or some intrinsic symmetry. Thus, some of the $L_{j}$'s may be of the same order
as $R$. 

To proceed we assume that only a finite number of terms in
expansion \eqref{general_action} survive at low energies and that all higher order terms are
suppressed. Therefore we must decide, in advance, a given order for
truncation. Once we have chosen such an order, we must include all
operators up to the corresponding mass dimension. All the coefficients are then low energy parameters to be fixed by experiment. Then we can go on to analyse the quantum dynamics of the fluctuations of the fields. When doing so, we have to keep in mind that \cite{Burgess:2007pt}:
\begin{enumerate}
\item{The high energy contributions to any observable, must be
  suitably regularised in a way that does not affect the low energy physics.}
\item{Any quantity which can be eliminated by a redefinition of the low
  energy parameters ($\alpha_jL_j^{p_j+q_j}$) must be discarded.}
\item{Any contribution which is of higher order in $L_{Pl}/R$ than the
chosen order for truncation, must be discarded.}
\end{enumerate}

\subsection{The action}\label{the_action}

Consider a scalar field $\Phi$, with mass
$\mu$ coupled only to gravity (i.e. with no self coupling or
equivalently potential), through the Ricci
scalar,  $\mathcal{R}$, of a curved space-time. The scalar field part of the theory is described by the action

\bequ \label{general_action2}
\mathcal{S}_{\Phi}=\int d^{n+1}x \sqrt{-g}\left(-\frac{1}{2}\partial_{\alpha}\Phi\partial^{\alpha}\Phi-\frac{1}{2}\mu^2\Phi^2-\frac{1}{2}\xi \mathcal{R}\Phi^2\right) \ . \eequ
These are all the possible operators  with mass
dimension up to $n+1$, which couple to gravity, contain no
self coupling and that will contribute in our model. Indeed, the only remaining possible operators are of the form  $\Phi(C+\mathcal{R}^{(p_j)})$, where $C$ is some constant. By a field redefinition these can be exchanged by operators of the form $\partial \Phi\partial \mathcal{R}^{(p_j)}$ and others which are purely gravitational. The latter may be absorbed in a redefinition of the gravitational couplings in \eqref{action3}; the former will not contribute to the scalar field equation of motion in our model, since all curvature invariants will be constants. We shall further assume that all other higher order operators are suppressed by powers of $L_{Pl}/R$ (which is equivalent to consider that all corresponding $L_j$'s are of order $L_{Pl}$). 

Analogously for pure gravity  we must consider  all
gravitational operators up to that same order of truncation $p_j+q_j=n+1$ (c.f. \eqref{general_action}). Therefore the action for gravity
is,
\bequ
\mathcal{S}_{g_{\mu\nu}}=\frac{1}{16\pi G_N^{n+1}}\int
d^{n+1}x\sqrt{-g}\left(-2\Lambda+\mathcal{R}+\sum_{p=2}^{[(n+1)/2]}\alpha^{(p)}\mathcal{R}^{(2p)}\right) \label{action3}
\ , \eequ
where $[\dots]$ denotes integer part. Note that all gravitational operators are of even dimension. On dimensional grounds, from
\eqref{general_action}, we expect all terms in (\ref{action3}) which are of quadratic (or higher) order in the curvature to be heavily suppressed
by powers of the Planck mass (if we assume a single fundamental scale
for gravity).\footnote{Moreover there are strong experimental constraints
 \cite{Birrel} on theories of gravity with higher order terms, which
show that the extra coefficients must be small.  We could also claim that
there is some symmetry in the fundamental theory which sets them
exactly to zero. Neither of these arguments is strictly
necessary since we expect these terms to be suppressed in the first
place.} However, for consistency, we still have to include them in our
effective theory, because they may get renormalised, even though we can 
neglect them for the purpose of looking at the evolution equations for
the radius of the Universe at scales where the effective theory holds.

Thus, the total action describing our effective theory is:
\bequ
\mathcal{S}=\mathcal{S}_{g_{\mu\nu}}+\mathcal{S}_{\Phi}+\mathcal{S}_{Matter} \ ,\label{totalaction} \eequ
where the last term represents the contribution of the classical
matter that will be considered to be a perfect fluid with positive energy
density and zero pressure. In this simplified model we are neglecting the
interaction of the scalar field with the classical matter.

\subsection{Bare energy density}
Now we look at the quantum fluctuations of the scalar field on the
background metric. The equation of motion for $\Phi$, obtained from \eqref{general_action2}, is
\[ \left(\Box -\xi \mathcal{R}\right)\Phi=\mu^2\Phi \ , \]
and its energy momentum tensor
\bequ
T_{\mu
  \nu}^{\Phi}=\partial_{\mu}\Phi\partial_{\nu}\Phi+\xi\left(\mathcal{R}_{\mu\nu}-D_{\mu}D_{\nu}\right)\Phi^2+g_{\mu\nu}\left(2\xi-\frac{1}{2}\right)\left[\partial_{\alpha}\Phi\partial^{\alpha}\Phi+(\mu^2+\xi\mathcal{R})\Phi^2\right] \ . \label{EM_tensor} \eequ
Conformal coupling is obtained  by taking the coefficient $\xi=(n-1)/4n$ (and the theory is then conformal if $\mu=0$ as can be checked by noting that $T_{\mu\nu}g^{\mu \nu}=0$), whereas minimal coupling corresponds to $\xi=0$. We will fix the space-time geometry to be the one of an $n+1$ dimensional Einstein Static Universe: 
\bequ
ds^2_{ESU^n}=-dt^2+R^2d\Omega_{S^n} \ , \label{metric}\eequ whose Ricci scalar is $\mathcal{R}=n(n-1)/R^2$, and where we have identified the radius of the universe with the scale $R$ introduced just before equation \eqref{general_action}. This is a solution of \eqref{totalaction}, if one takes $\Phi=0=\alpha^{(p)}$ and the cosmological constant and energy density of the perfect fluid to be related to $R$ by
\bequ
\rho=\frac{2\Lambda}{n-2}=\frac{n-1}{R^2} \ . \label{classicalsol}\eequ

The eigenfunctions of the Klein-Gordon operator $(\Box -\xi \mathcal{R}-\mu^2)$ may be taken in the form
\bequ\Phi^{(n)}_{\ell,\dots}= e^{-i\omega t}Y^{(n)}_{\ell,\dots}(\theta) \ , \label{eigenfunctions} \eequ
where the dots represent all necessary quantum numbers to completely characterise the hyper-spherical harmonic $Y^{(n)}_{\ell,\dots}(\theta)$ on $S^n$, with angular coordinates $\theta$ (see, for instance \cite{caillol,hyper_harmonics}). With this ansatz, we find the frequency spectrum 
\bequ \omega^{(n)}_{\ell}=\frac{\sqrt{\ell(\ell+n-1)+\xi n(n-1)+R^2\mu^2}}{R} \ , \ \ \ \ \ \ell\in \mathbb{N}_0 \ ,  \label{frespe}\eequ
with each frequency having degeneracy
\bequ d^{(n)}_{\ell}=\frac{(2\ell+n-1)\Gamma(n+\ell-1)}{\Gamma(n)\Gamma(\ell+1)} \ . \label{deg}\eequ
The unrenormalised vacuum energy density is obtained by a straightforward generalisation of the canonical quantisation procedure discussed in \cite{Herdeiro:2005zj}; using the notation therein, we obtain the expected result: 
\bequ
\rho_0^{(n)}\equiv \langle \,
\left(T^{\Phi}\right)^0_{\ 0}\,\rangle=\frac{1}{2V^{(n)}}\sum_{\ell=0}^{+\infty}d^{(n)}_{\ell}\omega^{(n)}_{\ell} \ ,
\label{bare_energy_density} \eequ
where $V^{(n)}$ is the volume (strictly speaking, the hyper-area) of $S^n$.

For future notational convenience, we define at this point
\bequ
k\equiv \frac{\ell+\delta}{R} \ ,\ \   a^2\equiv
\frac{\eta+\mu^{2}R^{2}}{R^2} \ ,\ \  \eta\equiv-\delta^2+\xi n(n-1)\
,\ \  \delta\equiv \frac{n-1}{2} \ ; \label{defi}\eequ
thus \eqref{frespe} may be rewritten
\begin{equation}
\omega^{(n)}_{\ell}=\sqrt{k^2+a^2} \ . 
\label{frespe2}
\end{equation}
Manipulating \eqref{deg} we can also rewrite it as
\begin{equation}
d^{(n)}_{\ell}=\dfrac{2kR}{(n-1)!}\prod_{i=0}^{n-3}\left[kR+\frac{n-3-2i}{2}\right]\equiv
P_{-}(kR)\ , \ \ \ \ \ n\ge 3 \ , \label{frespe21}
\end{equation}
where we have introduced the useful notation
\begin{equation}
P_{\pm}(\tau)\equiv \left\{ 
\begin{array}{c} 
\displaystyle{\dfrac{2}{(2p+2)!}\prod_{j=0}^{p}\left[\tau^2\pm
    j^2\right]= \tau^2Q_{\pm}(\tau^2) \ , \ \ \ n=2p+3}  \  , \\
\displaystyle{\dfrac{2\tau}{(2p+3)!}\prod_{j=0}^{p}\left[\tau^2\pm
    \left(j+\frac{1}{2}\right)^2\right]= \tau Q_{\pm}(\tau^2)\ , \ \ \  n=2p+4 \ , }
\end{array} 
\right. \label{Pt}
\end{equation}
and the expansion
\begin{equation}
Q_{\pm}(\tau^2)\equiv \left\{ 
\begin{array}{c} 
\displaystyle{\sum_{m=0}^{p}(\pm 1)^{p+1+m}\alpha_m \tau^{2m} \ , \ \ \ n=2p+3}  \  , \\
\displaystyle{\sum_{m=0}^{p+1}(\pm 1)^{p+1+m}\beta_m \tau^{2m}\ , \ \ \  n=2p+4 \ , }
\end{array} 
\right. \label{Qt}
\end{equation}
which defines the (positive) constants $\alpha_m$ and $\beta_m$. Note that the case $n=2$ is obtained with the product in \eqref{frespe21} replaced by $1$, whereas the case $n=1$ is non-degenerate.

\subsection{A generic class of regularisations}\label{generic_reg}
We now define a regularisation prescription for the energy
density based on very weak and physically well motivated assumptions. This will allow us to separate and analyse the
divergent and finite contributions. By using the
first law of thermodynamics we can then obtain the regularised
pressure and therefore the full regularised energy momentum
tensor. In section \ref{red}, all the divergences and finite contributions which
scale with the background radius $R$ in the
same way as the various gravitational terms will be subtracted,
 renormalising the corresponding couplings.

To regularise the energy density we need to suppress the high
energy modes. This is achieved by introducing a regulator, which should actually arise naturally from the fundamental theory by integrating out the massive degrees of freedom; but our ignorance about such theory implies that the way in which this regulator fundamentally arises is unknown. Besides suppressing the high energy (or momentum) contributions to physical observables, the regulator leads to a cutoff dependence in the bare couplings of our effective theory.   But if this theory is to give a good low energy description, the
infrared physics will be regulator independent, so such dependence is assumed to cancel when we  renormalise the bare couplings

We assume that, in the fundamental theory, there is a natural length
scale $\gamma L$ which acts as
regularisation parameter ($L$ is a typical length
scale in our model and $\gamma$ will turn out to be a very small dimensionless number). In addition we can introduce other dimensionless
parameters, $\gamma_i$, which could, for instance, control the form of the regulator before it is removed; alternatively they could be associated with ratios of other length scales in the problem. A realisation of the former possibility will be seen in appendix \ref{mcutoff}. Thus, on dimensional grounds, a general regularised energy density which could arise from the fundamental theory is
\bequ
\rho^{(n)}_{\gamma}=\frac{1}{2V^{(n)}}\sum_{\ell=0}^{+\infty}d^{(n)}_{\ell}\omega^{(n)}_{\ell}g(\gamma
L\omega^{(n)}_{\ell},\gamma_i)
\ . \label{firstreg}
\eequ
The physical
condition that the infrared physics should be regulator independent, amounts to say that the regularising function $g$, introduced in \eqref{firstreg}, stays close to 1 for a wide range of
energies around zero. This is equivalent to say that for sufficiently small energies
\[
\dfrac{\delta g}{g}\simeq \delta g \ll \dfrac{\delta \omega}{\omega}\Rightarrow \gamma L\omega
\partial_0g \ll 1 \ , 
\]
where $\delta$ denotes a variation with energy and $\partial_0$ a
differentiation with respect to the first argument of $g$. Hence, to a
very good approximation 
\[
g(\gamma L\omega^{(n)}_{\ell},\gamma_i)\simeq
g(0,\gamma_i)\ .
\]
Thus, an order 1 rescaling of
$\gamma$ will not affect the low energy physics; this will turn out to be
an important condition used in the isolation of the divergences,
which is not surprising since $\gamma$ is to some extent arbitrary and
any arbitrariness must be absorbed together with the divergences.

In what follows, any remaining dimensionless parameters $\gamma_i$ will not affect the reasoning so they are omitted; we therefore write
\bequ
\rho^{(n)}_{\gamma}=\frac{1}{2V^{(n)}}\sum_{\ell=0}^{+\infty}d^{(n)}_{\ell}\omega^{(n)}_{\ell}g(\gamma
L\omega^{(n)}_{\ell})
\ .\label{reged0}
\eequ
One way to interpret this formula is that only the energy spectrum is distorted. This is physically reasonable since the background is taken to be fixed and thus the scalar field could still
be expanded in hyper-spherical harmonics, so $\ell$ should still be a
good quantum number. This type of regularisation
is therefore consistent with spacetime symmetries. We will confirm it for the special case of point splitting
regularisation in appendix \ref{pointsplit}.

It might seem that \eqref{reged0} simply introduces  a damping
function. It is certainly true that any damping function
regularisation is of this form. But we will make the point that, if we
leave the dimensionless function $g$ unspecified - and we will obtain
the renormalised energy density without specifying $g$ -,
\eqref{reged0} effectively includes a much broader class of
regularisations. Indeed, as shown in appendix \ref{commonmethods} for
the special case of point-splitting, it will become clear that $g$
does not even have to be a \textit{damping} function; it might simply
oscillate with increasing $\omega^{(n)}_{\ell}$  without ever
decreasing its
modulus. 

Since we expect $\gamma L$ to be of the order of the Planck length and
$L$ is of the order of the typical length scales in our problem (which are either
$R$ or $\mu^{-1}$), $\gamma$ should indeed be very small, which
justifies taking $\gamma$ to zero to identify the divergences. In fact, we will not have to set
it to zero, since all cutoff dependent contributions will explicitly
take a form which is readily absorbed in low energy couplings. The
same will hold for all regularisations considered in appendix \ref{commonmethods}, independently of any other extra parameters $\gamma_i$ controlling the form of the regulator.

For the calculations that follow, \eqref{reged0} is conveniently written in terms of a function $f_{\gamma}(\ell+\delta)$:
\bequ
\rho^{(n)}_{\gamma}=\frac{1}{2V^{(n)}}\sum_{\ell=0}^{+\infty}d^{(n)}_{\ell}\omega^{(n)}_{\ell}g(\gamma
L\omega^{(n)}_{\ell})\equiv\frac{1}{RV^{(n)}}\sum_{\ell=0}^{+\infty}f_{\gamma}(\ell+\delta)
\ .\label{reged}
\eequ

\subsection{Separating contributions with the Abel-Plana formula}\label{abel_plana}
To separate the divergences from the finite part, we will use
the Abel-Plana summation formula\footnote{For a thorough review of this formula and its applications see \cite{Saharian:2007ph}.}
\[
\sum_{m=0}^{+\infty} G(m)=\int_0^{+\infty}G(t)dt+\frac{G(0)}{2}+i\int_0^{+\infty}\frac{G(it)-G(-it)}{e^{2\pi t}-1}dt \ , \]
in a way that generalises the procedure in \cite{Herdeiro:2005zj}. The only restrictions on $G(z)$ are that  it must i) be analytical on the right
complex semi-plane ($z \in \mathbb{C}: \Re(z)\geq 0$) and ii) obey
\[
\lim_{y\rightarrow\infty}e^{-2\pi y}\left|G(x\pm iy)\right|=0 \ .
\]
These are very weak requirements which are obeyed by most well behaved
functions. Note that, when applied to \eqref{reged}, the second condition requires only that the function $f_{\gamma}$ does not \textit{grow faster} than exponentially as we go to infinity along the
imaginary axis. This will enable us to obtain an answer which is independent
of the regularisation and therefore consistent with any reasonable
high energy theory which reduces to our effective theory.

Applying the Abel-Plana formula to \eqref{reged} gives (with a simple manipulation that will become useful)
\bequ
\barr{l}
\displaystyle{\rho^{(n)}_{\gamma}RV^{(n)}=\int_{t_{min}}^{+\infty}f_\gamma(t)dt-\int_{t_{min}}^{\delta}f_\gamma(t)dt+\frac{f_\gamma(\delta)}{2}+i\int_0^{+\infty}\frac{f_\gamma(it+\delta)-f_\gamma(-it+\delta)}{e^{2\pi t}-1}dt} \spa{0.4cm}\\ \displaystyle{~~~~~~~~~~~~~\equiv I_{\gamma,div}+I_{\gamma,fin}} \earr \label{reg_abelplana}\ , \eequ
where $t_{min}=0$ for $a^2\geq0$ and $t_{min}=|aR|$ otherwise; $I_{\gamma,div}$ is a divergent contribution (in the $\gamma\rightarrow 0$ limit) corresponding to the first integral and $I_{\gamma,fin}$ is a finite contribution corresponding to the last three terms. Note that the renormalisation procedure will \textit{not} simply amount to subtract $I_{\gamma,div}$, c.f. section \ref{red}. Using complex variable integration to simplify the latter (see appendix \ref{Iap}) yields
\begin{equation}
\lim_{\gamma\rightarrow 0}I_{\gamma,fin}=\left\{\begin{array}{ll}
\displaystyle{(-1)^p\int_{aR}^{+\infty}d\tau\dfrac{P_{+}(\tau)\sqrt{\tau^2-a^2R^2}}{\exp(2\pi\tau)-1}\
,}& n=2p+3 \ , \\
& \\
\displaystyle{(-1)^{p+1}\int_{0}^{aR}d\tau\dfrac{P_{+}(\tau)\sqrt{a^2R^2-\tau^2}}{\exp(2\pi\tau)+1}\ , }& n=2p+4 \ ,
\end{array}\right. \label{int1}
\end{equation}
for $a^2\geq0$ and
\begin{equation}
\lim_{\gamma\rightarrow 0}I_{\gamma,fin}=\left\{\begin{array}{ll}
\displaystyle{(-1)^p\int_{0}^{+\infty}d\tau\dfrac{P_{+}(\tau)\sqrt{\tau^2-a^2R^2}}{\exp(2\pi\tau)-1}+} &\\
\hspace*{7mm}\displaystyle{+\dfrac{1}{2}\int_{0}^{|a|R} dt P_-(t)\sqrt{-a^2R^2-t^2}\dfrac{\cos{\pi t}}{\sin{\pi t}}\ ,}& n=2p+3 \ , \\
& \\
\displaystyle{-\dfrac{1}{2}\int_{0}^{|a|R} dt P_-(t)\sqrt{-a^2R^2-t^2}\dfrac{\sin{\pi t}}{\cos{\pi t}}\ ,}& n=2p+4 \ ,\end{array}\label{int2}
\right.
\end{equation}
for $a^2<0$.  To deal with the divergent contribution start by performing the change of integration variable
\[
u\equiv R\omega^{(n)}_{t} \ , 
\]
to get
\[
I_{\gamma,div}= \left\{ 
\begin{array}{c} 
\displaystyle{\frac{1}{2}\int_{u_0}^{+\infty}du\,u^2\sqrt{u^2-a^2R^2}Q_{-}(u^2-a^2R^2)g\left(\frac{\gamma
  L}{R}
  u\right) \ , \ \ \ n=2p+3}  \  , \\
\displaystyle{\frac{1}{2}\int_{u_0}^{+\infty}du\,u^2Q_{-}(u^2-a^2R^2)g\left(\frac{\gamma
  L}{R}
  u\right)\ , \ \ \  n=2p+4 \ , }
\end{array}\right.
\]
where $u_0=aR$ for $a^2\geq 0$ and $u_0=0$ otherwise. It is now convenient to separate the analysis according to the parity of $n$.

\begin{description}
\item[$\bullet$] For $n=2p+3$ perform the change of variable 
\[
u^{2}=|a^2R^2|v^{2} \ , 
\]
to obtain
\[
I_{\gamma,div}=\frac{a^{4}R^4}{2}\int_{v_{0}}^{+\infty}dv\,\sqrt{v^{2}-s}Q_{-}(s(aR)^{2}(v^{2}-s))v^{2}g\left(\gamma|a|Lv\right) \ , 
\]
where $v_0=1$ for $a^2\geq 0$ and $v_0=0$ otherwise and $s$ is the sign of $a^2$. As argued in the last section, an order 1 rescaling of $\gamma$ should not affect the low energy physics; thus we have the freedom to perform the finite rescaling, $\tilde{\gamma}=|a|L\gamma$. Then we have 
\[
I_{\gamma,div}=\frac{a^{4}R^4}{2}\int_{v_0}^{+\infty}dv\,\sqrt{v^{2}-s}Q_{-}(s(aR)^{2}(v^{2}-s))v^{2}g\left(\tilde{\gamma}v\right)\ .
\]
Thus, the divergent contribution in \eqref{reg_abelplana} is obtained by putting all the pre-factors back and using the expansion \eqref{Qt}:
\begin{equation}
\rho_{\tilde{\gamma},div}=\sum_{q=0}^{p+2}\frac{A_{q,\mu,\tilde{\gamma},\xi,p}}{R^{2q}} \ , \label{rr1}
\end{equation}
where 
\begin{multline}
A_{q,\mu,\tilde{\gamma},\xi,p}\equiv\sum_{m=m(q)}^{p}\dfrac{(-1)^{p+1+m}s^m\alpha_m}{2V^{(n)}_{R=1}}\binom{m+2}{p+2-q}\mu^{2(p+2-q)}\eta^{m-p+q}\times
\\ \times\int_{v_0}^{+\infty}dv\,
v^{2}(v^{2}-s)^{m+1/2}g\left(\tilde{\gamma}v\right) \ ; \nonumber
\end{multline}
$\eta$ is defined by \eqref{defi} and $m(q)=0$ for $q={p+1,p+2}$ and $m(q)=p-q$ otherwise. It already becomes clear from \eqref{rr1} that all divergent contributions scale with $R$ in the same way as the gravitational operators in \eqref{action3}.

\item[$\bullet$] For $n=2p+4$ we simply perform the change of variable $uL/R\equiv v$; extend the integration to $v=0$ and subtract the corresponding term, which exists only for $a^2\ge 0$. Using the expansion for $Q_{-}$ we get,
\begin{equation}
\rho_{{\gamma},div}=\sum_{q=0}^{p+1}\frac{B_{q,\mu,\gamma,\xi,p}}{R^{2q}} \ , \label{rr2}
\end{equation}
where 
\begin{multline}
B_{q,\mu,\gamma,\xi,p}\equiv\sum_{k=0}^{q}\dfrac{(-1)^{q}\beta_{p+1-q+k}}{2V^{(n)}_{R=1}}\binom{p+1-q+k}{k}\dfrac{\eta^{k}}{L^{2p-2q+5}}\times
\\ \times\int_{0}^{+\infty}dv\,
v^{2}(v^{2}-\mu^2L^2)^{p+1-q}g\left(\gamma v\right) \ , \nonumber
\end{multline}
and an extra finite term which is only present for $a^2\geq0$,
\begin{equation}
  \rho_{extra}=\sum_{m=0}^{p+1}\dfrac{b_m(aR)^{2m+3}}{R^{2p+5}} \ , \label{rhoextra}
\end{equation}
where we have defined the coefficients $b_m$ as
\[
b_m\equiv (-1)^p\frac{ 2^{m-1}m!}{(3+2m)!!\, V^{(n)}_{R=1}}\beta_m \ , 
\]
and where $V_{R=1}^{(n)}$ is the volume of the $n$-sphere with unit radius. 
\end{description}

Assembling the results \eqref{int1}, \eqref{int2}-\eqref{rhoextra}, we finally get the full regularised result for \eqref{reg_abelplana}, up to order $\tilde{\gamma}^0$ ($\gamma^0$):
\begin{equation}
\rho^{(n)}_{reg}=\left\{\begin{array}{ll}
\displaystyle{\sum_{q=0}^{p+2}\frac{A_{q,\mu,\tilde{\gamma},\xi,p}}{R^{2q}}+\dfrac{(-1)^p}{RV^{(n)}}\int_{aR}^{+\infty}d\tau\dfrac{P_{+}(\tau)\sqrt{\tau^2-a^2R^2}}{\exp(2\pi\tau)-1}\
,}& n=2p+3 \ , \\
& \\
\displaystyle{\sum_{q=0}^{p+1}\dfrac{B_{q,\mu,\gamma,\xi,p}}{R^{2q}}+\sum_{m=0}^{p+1}\dfrac{b_m(aR)^{2m+3}}{R^{2p+5}}+\dfrac{(-1)^{p+1}}{RV^{(n)}}\int_{0}^{aR}d\tau\dfrac{P_{+}(\tau)\sqrt{a^2R^2-\tau^2}}{\exp(2\pi\tau)+1} \ , }& n=2p+4 \ , 
\end{array}\right.\label{reg_rho1} 
\end{equation}
for $a^2\geq0$ and
\begin{equation}
\rho^{(n)}_{reg}=\left\{\begin{array}{ll}
\displaystyle{\sum_{q=0}^{p+2}\frac{A_{q,\mu,\tilde{\gamma},\xi,p}}{R^{2q}}+\dfrac{(-1)^p}{RV^{(n)}}\int_{0}^{+\infty}d\tau\dfrac{P_{+}(\tau)\sqrt{\tau^2-a^2R^2}}{\exp(2\pi\tau)-1}+} &\\
\hspace*{7mm}\displaystyle{+\dfrac{1}{2RV^{(n)}}\int_{0}^{|a|R} dt P_-(t)\sqrt{-a^2R^2-t^2}\dfrac{\cos{\pi t}}{\sin{\pi t}}\ ,}& n=2p+3 \ , \\
& \\
\displaystyle{\sum_{q=0}^{p+1}\dfrac{B_{q,\mu,\gamma,\xi,p}}{R^{2q}}-\dfrac{1}{2RV^{(n)}}\int_{0}^{|a|R} dt P_-(t)\sqrt{-a^2R^2-t^2}\dfrac{\sin{\pi t}}{\cos{\pi t}}\ ,}& n=2p+4 \ ,\label{reg_rho2}\end{array}
\right.
\end{equation}
for $a^2<0$. Now that we have the regularised energy density, the
regularised pressure can be easily obtained via the first law of
thermodynamics, as in \cite{Herdeiro:2005zj}, 
\begin{equation}
p^{(n)}=-\frac{\partial(\rho^{(n)}R^n)}{\partial R^n} \ . 
\label{firstlaw}
\end{equation}
This is only a statement of conservation of energy (or equivalently
covariant conservation of the energy momentum tensor) which is one of the
good properties of our low energy theory that we want to keep. The
result is
\begin{equation}
p^{(n)}_{reg}=\left\{\begin{array}{ll}
\displaystyle{-\sum_{q=0}^{p+2}\frac{(n-2q)A_{q,\mu,\tilde{\gamma},\xi,p}}{nR^{2q}}+\dfrac{(-1)^p}{nRV^{(n)}}\int_{aR}^{+\infty}d\tau\dfrac{P_{+}(\tau)(\tau^2-a^2R^2+\mu^2R^2)}{\sqrt{\tau^2-a^2R^2}(\exp(2\pi\tau)-1)}\
,}& n=2p+3 \ , \\
& \\
\displaystyle{-\sum_{q=0}^{p+1}\dfrac{(n-2q)B_{q,\mu,\gamma,\xi,p}}{nR^{2q}}+\sum_{m=0}^{p+1}\dfrac{b_m\left[(aR)^{2m+3}-(2m+3)(aR)^{2m+1}\mu^2R^2\right]}{nR^{2p+5}}+}
& \\ \hspace*{40mm}
\displaystyle{+\dfrac{(-1)^{p+1}}{nRV^{(n)}}\int_{0}^{aR}d\tau\dfrac{P_{+}(\tau)(a^2R^2-\mu^2R^2-\tau^2)}{\sqrt{a^2R^2-\tau^2}(\exp(2\pi\tau)+1)}\ ,}
& n=2p+4 \ ,
\end{array}\right. \label{reg_p1}
\end{equation}
for $a^2\geq0$ and
\begin{equation}
p^{(n)}_{reg}=\left\{\begin{array}{ll}
\displaystyle{-\sum_{q=0}^{p+2}\frac{(n-2q)A_{q,\mu,\tilde{\gamma},\xi,p}}{nR^{2q}}+\dfrac{(-1)^p}{nRV^{(n)}}\int_{0}^{+\infty}d\tau\dfrac{P_{+}(\tau)(\tau^2-a^2R^2+\mu^2R^2)}{\sqrt{\tau^2-a^2R^2}(\exp(2\pi\tau)-1)}+} &\\
\hspace*{40mm}\displaystyle{+\dfrac{1}{2nRV^{(n)}}\int_{0}^{|a|R} dt P_-(t)\dfrac{\mu^2R^2-a^2R^2-t^2}{\sqrt{-a^2R^2-t^2}}\dfrac{\cos{\pi t}}{\sin{\pi t}}\ ,}& n=2p+3 \ , \\
& \\
\displaystyle{-\sum_{q=0}^{p+1}\dfrac{(n-2q)B_{q,\mu,\gamma,\xi,p}}{nR^{2q}}-\dfrac{1}{2nRV^{(n)}}\int_{0}^{|a|R} dt P_-(t)\dfrac{\mu^2R^2-a^2R^2-t^2}{\sqrt{-a^2R^2-t^2}}\dfrac{\sin{\pi t}}{\cos{\pi t}}\ ,}& n=2p+4 \ ,\end{array}
\right.\label{reg_p2}
\end{equation}
for $a^2<0$.

\subsection{Renormalising the energy-momentum tensor}
\label{red}
Now that we have separated all the divergences from the finite result,
we proceed to renormalise the energy momentum tensor. To do it, any high energy contributions to the physical observables must be absorbed into the set of gravitational
couplings in (\ref{action3}), which are fixed at low energies. Such contributions are the divergences that we have identified in
\eqref{reg_rho1}-\eqref{reg_p2}; thus we need to show that they can be absorbed into the aforementioned gravitational couplings. 

Varying the action \eqref{totalaction} we find the Einstein equations
\[
G_{\mu\nu}+\Lambda g_{\mu\nu}+\sum_{p=2}^{[(n+1)/2]}\alpha^{(p)}\left[\frac{\partial\mathcal{R}^{(2p)}}{\partial g^{\mu\nu}}-\frac{1}{2}g_{\mu\nu}\mathcal{R}^{(2p)}\right]=8\pi G_N^{n+1}\left(T_{\mu\nu}^{\Phi}+T_{\mu\nu}^{Matter}\right) \ , \]
where the only contribution to the scalar field energy momentum tensor is its vev $\langle T^{\Phi}_{\mu \nu}\rangle$. Due to the maximal symmetry of the spatial sections of the background \eqref{metric}, these equations reduce to a generalised Friedmann equation
\bequ
-\Lambda+\frac{n(n-1)}{2}\frac{\dot{R}^2+1}{R^2}+\sum_{q=2}^{[(n+1)/2]}\alpha^{(q)}\frac{F^{(q)}(n,\dot{R})}{R^{2q}}=\kappa^2\left[\rho^{(n)}_{cla}+\rho^{(n)}_{ren}\right] \ , \label{cosmo12}\eequ
where the explicit form of the functions $F^{(q)}(n,\dot{R})$ depends on the curvature invariants $\mathcal{R}^{(2q)}$, and to a generalised Raychaudhuri equation
\bequ
n(n-1)\frac{\ddot{R}(t)}{R(t)}-2\Lambda-\left(\frac{n-2}{R^n}-\frac{R^{1-n}}{\dot{R}}\frac{\partial}{\partial t}\right)\sum_{q=2}^{[(n+1)/2]}\alpha^{(q)}\frac{F^{(q)}(n,\dot{R})}{R^{2q}}=-\kappa^2(n-2)\left[\sigma^{(n)}_{cla}+\sigma^{(n)}_{ren}\right] \ . \label{cosmo22}\eequ
The latter equation can be obtained from \eqref{cosmo12} together with \eqref{firstlaw}. Hence, to prove that we can renormalise all divergences it suffices to consider the Friedmann equation \eqref{cosmo12}. It is then manifest that the scaling with $R$ of the divergences shown in \eqref{rr1} and \eqref{rr2} is the correct one 
to be absorbed in a renormalisation of the cosmological constant ($q=0$), Newton constant ($q=1$) and gravitational couplings $\alpha^{(q)}$ ($q>1$).

It is interesting to note that, by using a very generic regulator, we
obtained almost all possible divergences. The single exception arises for 
even spheres for which the gravitational coupling of order $q=2p+4$ does not get renormalised (by a divergent contribution). Which divergences show up depends
on the specific regularisation method and hence on the high energy completion of the theory. 

Also interesting is the fact that we can accommodate theories with special symmetries which forbid some higher curvature terms in the gravitational Lagrangian, by choosing regulators which integrate some of the divergences to zero (or very close). Thus, this treatment is consistent, for example, with a theory of gravity
which contains only the usual Einstein action. The fact that all the
other couplings are small may be seen as an intrinsic property of the
theory at the given energy scale.

After this subtraction, we are left with a finite result. However, we
must check if there are any remaining \textit{finite} ambiguities. These would be additive contributions which are
of the form $~R^{-2q}$ with a pre-factor independent of $R$. Since these could be absorbed in a renormalisation of the gravitational couplings they would lead to an ambiguity.   The easiest way to look for such terms is to consider the asymptotic expansion of the energy density in the limit when $R\rightarrow \infty$. Equation \eqref{firstlaw}  can then be used to obtain the asymptotic expansion for
the pressure. With these expansions we easily identify the various subtractions that must be made, by looking at powers of $R^{-1}$. 

For the case of odd spheres ($n=2p+3$), the dominant contribution in the finite part of the energy density \eqref{reg_rho1} is, in the large $R$ limit,  
\bequ
\rho^{(2p+3)}_{reg,fin}=(-1)^p\frac{(\mu R)^{2p+5/2}}{2\pi(2p+2)!RV^{(n)}}e^{-2\pi \mu R}+\dots \ , \label{final1gg}\eequ
where `$\dots$' denote sub-dominant terms; hence, for odd spheres, the dominating term is exponentially damped and no extra renormalisation is needed. The corresponding terms for even spheres ($n=2p+4$) come  from the two finite terms in \eqref{reg_rho1}:
\[
\barr{l}
\displaystyle{\rho^{(2p+4)}_{reg,fin}=\sum_{m=0}^{p+1}\sum_{k=0}^{m+1}\frac{b_m\eta^k\mu^{2m-2k+3}}{R^{2(p+k-m+1)}}\binom{m+3/2}{k}~~~~~~~~~~~~~~~}\spa{0.3cm}\\
\displaystyle{~~~~~~~~~~+\sum_{k=0}^{p+1}\beta_k(-1)^{p+k}\zeta_{R}(-2k-1)\left(\frac{1}{2}-\frac{1}{2^{2k+2}}\right)\frac{\mu}{V^{(n)}_{R=1}R^{2p+4}}+\dots \equiv \rho^{(2p+4)}_{extra}+\dots \ ,} \earr \]
where $\beta_k$ were defined in \eqref{Qt}; hence, for even spheres, we need to subtract $\rho^{(2p+4)}_{extra}$ in order to get the final renormalised result. To obtain the term to be subtracted to the pressure we use \eqref{firstlaw}, which yields
\[
p^{(2p+4)}_{extra}\equiv \sum_{m=0}^{p+1}\sum_{k=0}^{m+1}\frac{2(k-m-1)b_m\eta^k\mu^{2m-2k+3}}{(2p+4)R^{2(p+k-m+1)}}\binom{m+3/2}{k} \ . \]

A different way to argue that these subtractions are to be expected is as follows. On general grounds one expects the Casimir effect to be suppressed for
very massive fields. Indeed, the mass introduces an effective infrared cut-off. Since the Casimir force results from the infrared behaviour of quantum fluctuations in some confined setup, one expects the Casimir force to vanish as the
mass increases. One can check from \eqref{final1gg} that this is
indeed what happens for odd spheres.  However the terms in $\rho^{(2p+4)}_{extra}$ yield a contribution for the energy  that increases without bound for increasing mass. Similar terms are discussed in
\cite{review} (section 2.3 and 3.4) for the special case of the
2-sphere. Therein it is argued that an extra renormalisation must be
imposed which eliminates them. In our setup such extra renormalisation appears naturally.

Finally we can exhibit the  renormalised energy density and pressure. For $n$ odd they are
\begin{equation}
\begin{array}{ll}
\rho^{(2p+3)}_{ren}=\\ \left\{\begin{array}{ll}
\displaystyle{\dfrac{(-1)^p}{RV^{(n)}}\int_{aR}^{+\infty}d\tau\dfrac{P_{+}(\tau)\sqrt{\tau^2-a^2R^2}}{\exp(2\pi\tau)-1}\
  ,}& (a^2\ge 0) \ , \\
& \\ \vspace{2mm}
\displaystyle{\dfrac{(-1)^p}{RV^{(n)}}\int_{0}^{+\infty}d\tau\dfrac{P_{+}(\tau)\sqrt{\tau^2-a^2R^2}}{\exp(2\pi\tau)-1}+\dfrac{1}{2RV^{(n)}}\int_{0}^{|a|R} dt
  P_-(t)\sqrt{-a^2R^2-t^2}\dfrac{\cos{\pi t}}{\sin{\pi t}}\ ,}& (a^2< 0) \ , 
  \end{array} \right.  \end{array}  \label{rho_ren1}
\end{equation}
\begin{equation}
p^{(2p+3)}_{ren}=\left\{\begin{array}{ll}
\displaystyle{\dfrac{(-1)^p}{nRV^{(n)}}\int_{aR}^{+\infty}d\tau\dfrac{P_{+}(\tau)(\tau^2-a^2R^2+\mu^2R^2)}{\sqrt{\tau^2-a^2R^2}(\exp(2\pi\tau)-1)}\
,}& (a^2\ge 0) \ , \\
& \\ \vspace{2mm}
\displaystyle{\dfrac{(-1)^p}{nRV^{(n)}}\int_{0}^{+\infty}d\tau\dfrac{P_{+}(\tau)(\tau^2-a^2R^2+\mu^2R^2)}{\sqrt{\tau^2-a^2R^2}(\exp(2\pi\tau)-1)}+} &\\
\hspace*{10mm}\displaystyle{+\dfrac{1}{2nRV^{(n)}}\int_{0}^{|a|R} dt
  P_-(t)\dfrac{\mu^2R^2-a^2R^2-t^2}{\sqrt{-a^2R^2-t^2}}\dfrac{\cos{\pi
      t}}{\sin{\pi t}}\ ,}& (a^2< 0) \ . \end{array} \right. \label{p_ren1}
\end{equation}
For even spheres the renormalised energy density and pressure are
\begin{equation}
\begin{array}{l}
\rho^{(2p+4)}_{ren}=\left\{\begin{array}{ll}
\displaystyle{\sum_{m=0}^{p+1}\dfrac{b_m(aR)^{2m+3}}{R^{2p+5}}+\dfrac{(-1)^{p+1}}{RV^{(n)}}\int_{0}^{aR}d\tau\dfrac{P_{+}(\tau)\sqrt{a^2R^2-\tau^2}}{\exp(2\pi\tau)+1}}\ , &
  (a^2\ge 0) \ , \\
 \displaystyle{-\dfrac{1}{2RV^{(n)}}\int_{0}^{|a|R} dt P_-(t)\sqrt{-a^2R^2-t^2}\dfrac{\sin{\pi t}}{\cos{\pi t}}\ ,}& (a^2<0) \ , 
\end{array}\right\} - \rho_{extra}^{(2p+4)}
\ ,
\end{array} \label{rho_ren2}
\end{equation}
\begin{equation}
\begin{array}{l}
p^{(2p+4)}_{ren}=\left\{\begin{array}{ll}
\displaystyle{\sum_{m=0}^{p+1}\dfrac{b_m\left((aR)^{2m+3}-(2m+3)(aR)^{2m+1}\mu^2R^2\right)}{nR^{2p+5}}+}
& \\ \hspace*{10mm}
\displaystyle{+\dfrac{(-1)^{p+1}}{nRV^{(n)}}\int_{0}^{aR}d\tau\dfrac{P_{+}(\tau)(a^2R^2-\mu^2R^2-\tau^2)}{\sqrt{a^2R^2-\tau^2}(\exp(2\pi\tau)+1)}\ ,}&
  (a^2\ge 0) \ , \vspace{2mm}\\ 
 \displaystyle{-\dfrac{1}{2nRV^{(n)}}\int_{0}^{|a|R} dt P_-(t)\dfrac{\mu^2R^2-a^2R^2-t^2}{\sqrt{-a^2R^2-t^2}}\dfrac{\sin{\pi t}}{\cos{\pi t}}\ ,}& (a^2<0) \ , 
\end{array}\right\} -p^{(2p+4)}_{extra}
    \ .
\end{array} \label{p_ren2}
\end{equation}

Note that even though we have defined the renormalised energy density
using two different branches, the prescription for regularisation and
subtraction of the divergences was exactly the same on
both sides. Thus we have subtracted the same finite term for both
branches in the renormalised quantities.  The fact that we have to divide in two branches comes only from mathematical manipulations; it is a consequence of our way of representing the finite contributions.

\section{Analysis of the renormalised vacuum energy momentum tensor}
\label{analysis}
Expressions \eqref{rho_ren1}-\eqref{p_ren2} are, generically, non-trivial functions of one discrete  and three continuous parameters ($n,\xi,\mu,R$). In this section we will analyse in detail the behaviour of $\rho^{(n)}_{ren}$ and $\sigma^{(n)}_{ren}$ -which play important roles in equations (\ref{cosmo1}) and (\ref{cosmo2})- resorting to some analytical behaviours and numerical plots. 

A first observation to simplify the analysis is that the $\mu$ dependence is quite simple. Indeed, it follows from \eqref{rho_ren1}-\eqref{p_ren2}, together with \eqref{sigma} that $\rho^{(n)}_{ren}$ and $\sigma^{(n)}_{ren}$ can be written in the form
\bequ
\rho^{(n)}_{ren}=\frac{\alpha^{(n)}_{ren}(\xi,\mu R)}{R^{n+1}} \ , \ \ \ \ \ \sigma^{(n)}_{ren}=\frac{\varsigma^{(n)}_{ren}(\xi,\mu R)}{R^{n+1}} \ ; \label{newsigmarho}\eequ
indeed this follows from dimensional analysis. Changing variable to $\bar{R}\equiv \mu R$, we have 
\[
\rho^{(n)}_{ren}=\mu^{n+1}\frac{\alpha^{(n)}_{ren}(\xi,\bar{R})}{\bar{R}^{n+1}} \ , \ \ \ \ \ \sigma^{(n)}_{ren}=\mu^{n+1}\frac{\varsigma^{(n)}_{ren}(\xi,\bar{R})}{\bar{R}^{n+1}} \ . \]
Thus, we can simply consider the behaviour of $\rho^{(n)}_{ren}$ and $\sigma^{(n)}_{ren}$ for $\mu=1$,\footnote{Note that this numerical value for $\mu$ has  dimension of inverse length, since we are using units with $\hbar=c=1$; re-introducing these constants the numerical value of the mass is $\mu\hbar/c$.} which will be done separately for even and odd $n$ in the next two subsections. The behaviour for any $\mu\neq 0$ is simply obtained multiplying the vertical axis by a factor of $\mu^{n+1}$ and dividing the horizontal axis by $\mu$. 

\begin{figure}[h!]
\begin{picture}(0,0)(0,0)
\end{picture}
\centering\epsfig{file=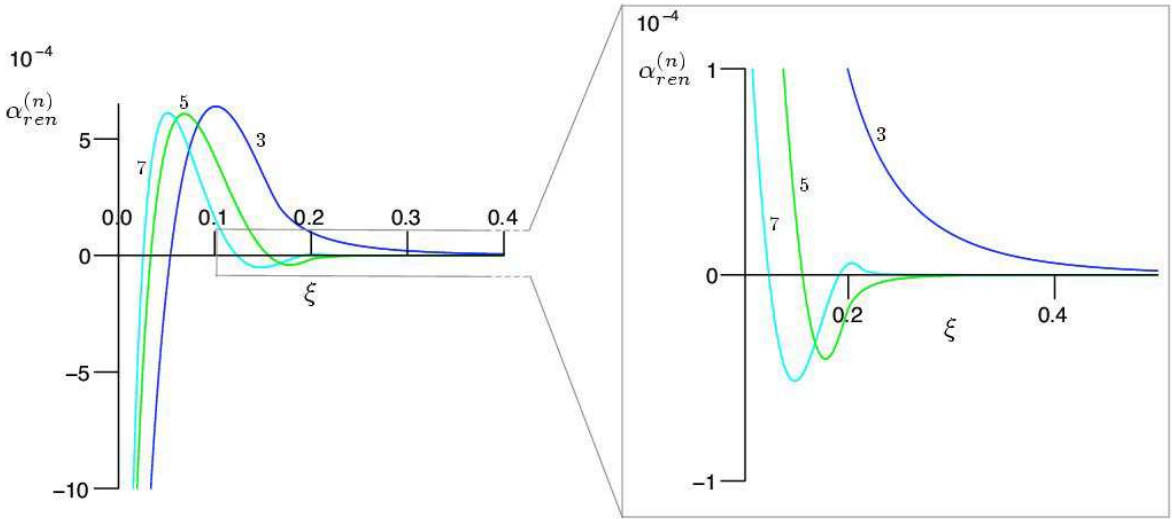,width=16cm}
\caption{$\alpha^{(n)}_{ren}$ versus $\xi$ for $\mu=0$ and $n=3$ (one zero), $n=5$ (two zeros) and $n=7$ (three zeros). Note that the sign oscillates with $n$ for conformal coupling.}
\label{alpha}
\end{figure}

\begin{figure}[h!]
\begin{picture}(0,0)(0,0)
\end{picture}
\centering\epsfig{file=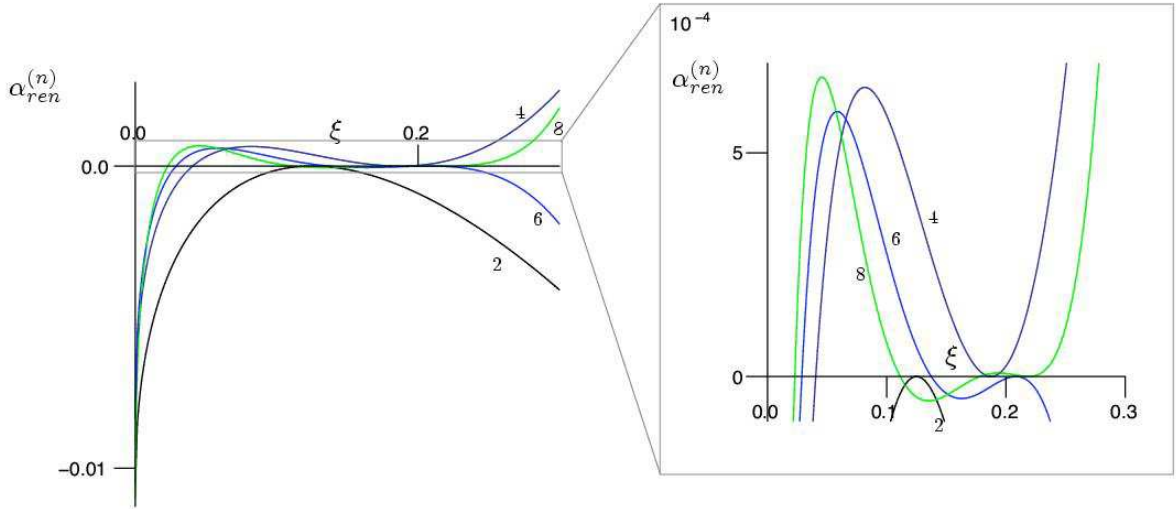,width=16cm} 
\caption{$\alpha^{(n)}_{ren}$ versus $\xi$ for $\mu=0$ and $n=2$ (one zero), $n=4$ (two zeros) $n=6$ (three zeros) and $n=8$ (four zeros). Note that one of the zeros always corresponds to an extreme, which occurs for conformal coupling.}
\label{rho2468}
\end{figure}

The massless case has to be considered separately, but the radial dependence is  very simple in such case. One can use \eqref{sigma}, \eqref{firstlaw} and \eqref{newsigmarho} to show that $p^{(n)}_{ren}=\rho^{(n)}_{ren}/n$ and hence\footnote{For $n=2$ the role of $\sigma$ is played by $\tilde{\sigma}^{(2)}\equiv 2 p^{(2)}$, which, in the massless case is $\tilde{\sigma}^{(2)}= \rho^{(2)}$.}
\bequ
\sigma^{(n)}_{ren}=\frac{n-1}{n-2}\rho^{(n)}_{ren} \ . \label{sigmamass0}\eequ
Thus $\sigma^{(n)}_{ren}$ is simply proportional to $\rho^{(n)}_{ren}$. The latter function has a very simple $R$ dependence; from \eqref{newsigmarho}, 
\bequ
\rho^{(n)}_{ren}=\frac{\alpha^{(n)}_{ren}(\xi)}{R^{n+1}} \ . \label{rhomass0}\eequ
Thus we just have to understand the behaviour of the function $\alpha^{(n)}_{ren}(\xi)$, which is displayed in figures \ref{alpha}  and \ref{rho2468} for the first few cases with $n$ odd and $n$ even respectively. Some generic features are: $\alpha^{(n)}_{ren}(\xi)$ is always negative for small $\xi$. For $n$ odd $\alpha^{(n)}_{ren}(\xi)$ has $(n-1)/2$ zeros and goes to zero for large $\xi$ with sign $(-1)^{(n+1)/2}$. For $n$ even $\alpha^{(n)}_{ren}(\xi)$ has $n/2$ zeros and diverges - as can be checked from \eqref{rho_ren2} - for large $\xi$ with sign $(-1)^{n/2}$ (since one of the zeros is an extreme).

A further special case is when $\xi= \xi_c\equiv (n-1)/4n$ (besides $\mu=0$), which corresponds to the conformal coupling. Then
\bequ
\alpha^{(n)}_{ren}(\xi_c)=\left\{\barr{cl}\displaystyle{\frac{(-1)^p}{V^{(n)}_{R=1}}\int_{0}^{+\infty}d\tau\dfrac{P_{+}(\tau)\tau}{\exp(2\pi\tau)-1}} & , \ n=2p+3  \spa{0.3cm}\\ 0& , \ n=2p+4 \earr \right.\ . \label{rhooddconf}
\eequ
The vacuum energy \textit{vanishes} for even dimensional spheres. For
odd dimensional spheres the vacuum energy approaches a finite non-zero
value whose sign \textit{alternates} with $n$, starting with a positive sign for $S^3$ (it is indeed negative for $S^1$ and zero for $S^2$ \cite{review}). These behaviours for the conformal case were first observed in \cite{ozcan}, and can be confirmed in  figures \ref{alpha}  and \ref{rho2468}, for the cases displayed therein. In particular, for the even $n$ case, the zero at $\xi_c$ is the one corresponding to an extreme in each curve.

Before proceeding to  a numerical analysis of  the behaviour of $\rho^{(n)}_{ren}$ and $\sigma^{(n)}_{ren}$ for $\mu\neq 0$,  let us make a final observation concerning the $\mu=0$ case. Since the radial dependence of the renormalised energy density is now of the type $1/R^{n+1}$ one could be led to think that such terms should be absorbed in the renormalisation of some gravitational coupling, following the discussion in section \ref{red}. Note, however, that the renormalisation process was appropriately implemented considering $\mu\neq 0$, since we must consider all operators with mass dimension up to $n+1$. In this case no further renormalisation was necessary. Once the final result was obtained one can study limits thereof (like the massless limit) without having to worry about further renormalisations.

\subsection{Odd $n$}

\begin{figure}[h!]
\begin{picture}(0,0)(0,0)
\end{picture}
\centering\epsfig{file=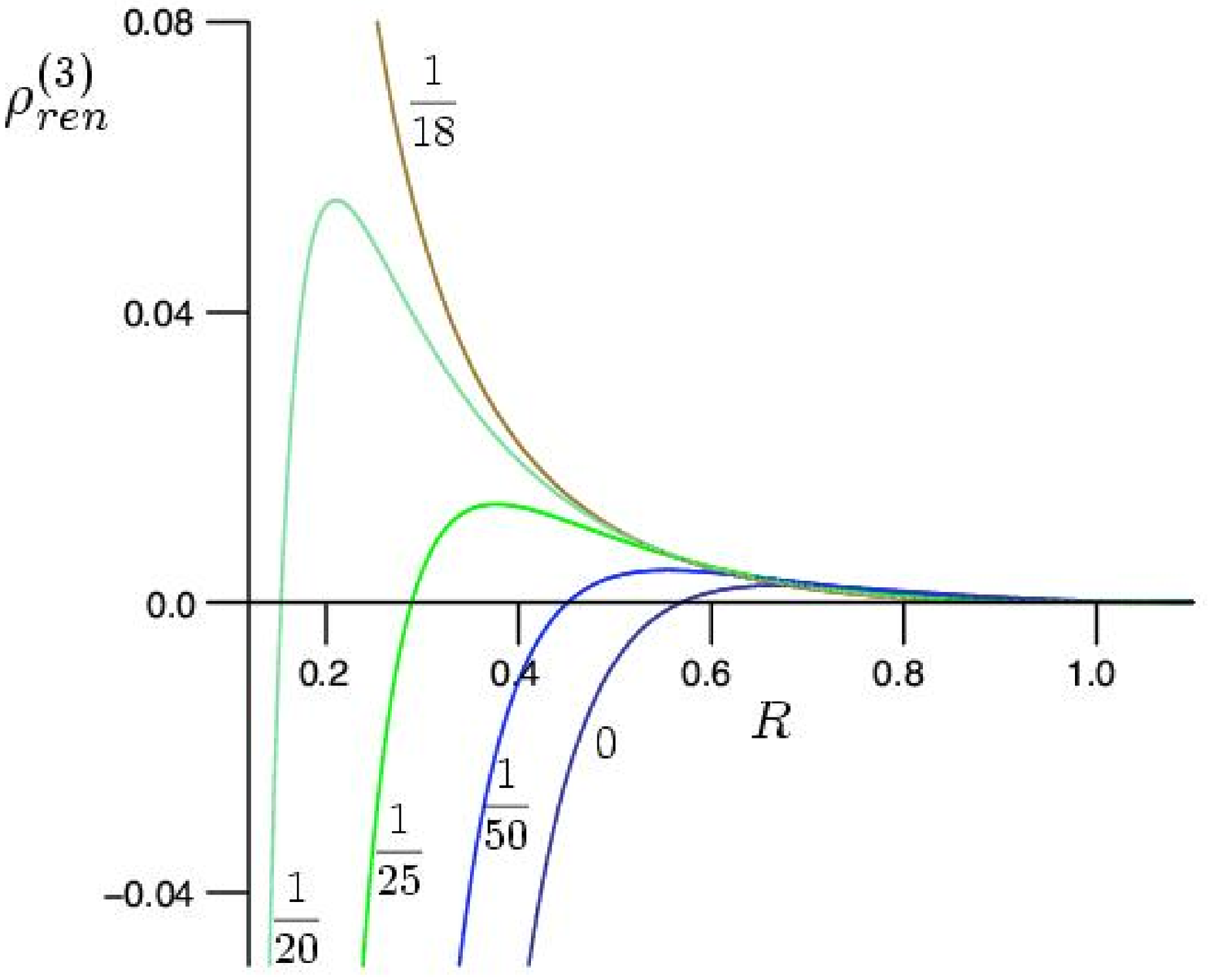,width=7.0cm} 
\centering\epsfig{file=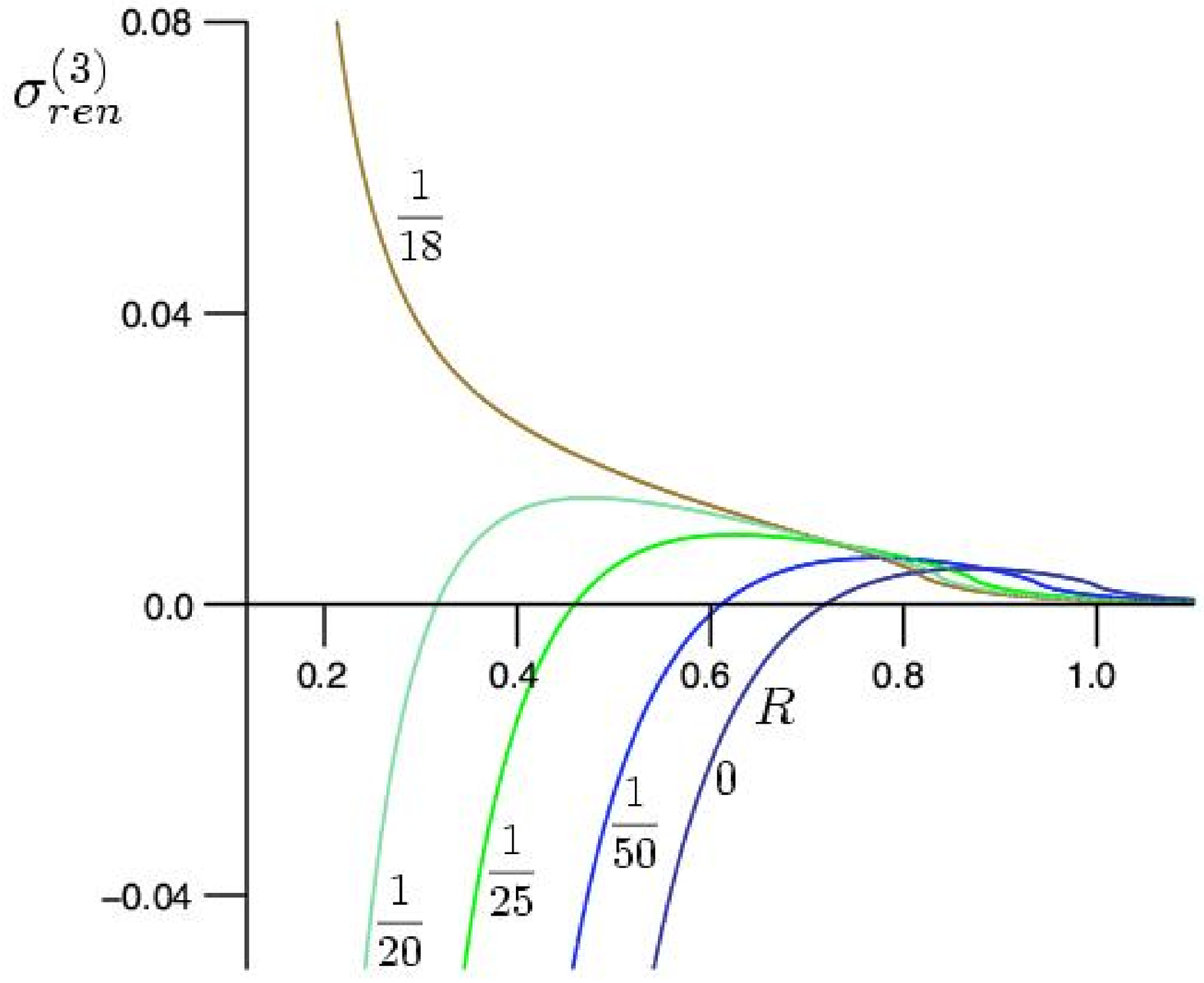,width=7.0cm}
\caption{$\rho^{(3)}_{ren}$ and $\sigma^{(3)}_{ren}$ versus $R$ for $\mu=1$ and $\xi=0,\frac{1}{50},\frac{1}{25},\frac{1}{20},\frac{1}{18}$, which are displayed next to the corresponding curve.}
\label{3esfera}
\end{figure} 

The first non-trivial case is $n=3$, for which $\rho^{(3)}_{ren}$ and $\sigma^{(3)}_{ren}$ are plotted in figure \ref{3esfera}, as functions of $R$, and for a number of representative couplings $\xi$. The behaviour of both functions is qualitatively similar, and can be summarised as follows. For large radius  $\rho^{(3)}_{ren}$ goes to zero through positive values, in accordance with \eqref{final1gg}. For sufficiently small coupling one verifies the existence of one zero. The $R$ value of the zero decreases with increasing coupling and, for sufficiently large coupling, the zero disappears and the curves become always positive. A consistency check is to verify \eqref{rhooddconf}; in this case, it implies that we must have $\rho_{ren}^{(3)}>0$  for $a=0$. Using \eqref{defi} for this case we verify that $a=0$ corresponds to $R=\sqrt{1-6\xi}$. One can then verify that for all couplings represented,  $\rho_{ren}^{(3)}$ is indeed positive for the corresponding $R$. Note also that t
 he zero disappears before we get to conformal coupling. For coupling larger than conformal ($\xi=1/6$ for $n=3$), only the positive branches of \eqref{rho_ren1} and \eqref{p_ren1} are needed to describe the behaviour of the renormalised quantities.

\begin{figure}[h!]
\begin{picture}(0,0)(0,0)
\end{picture}
\centering\epsfig{file=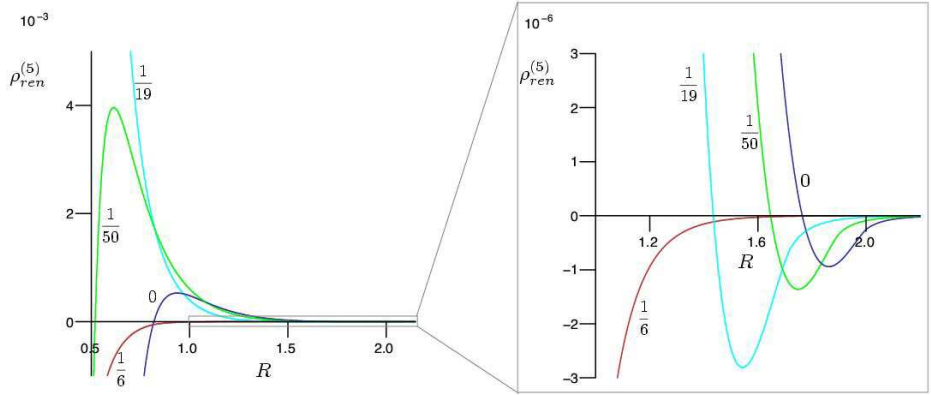,width=13cm}
\caption{$\rho^{(5)}_{ren}$ versus $R$ for $\mu=1$ and $\xi=0, \frac{1}{50}, \frac{1}{19}, \frac{1}{6}$.  In this and similar figures, the rhs figure gives a detail of the lhs one.}
\label{5esferarho}
\end{figure} 

\begin{figure}[b!]
\begin{picture}(0,0)(0,0)
\end{picture}
\centering\epsfig{file=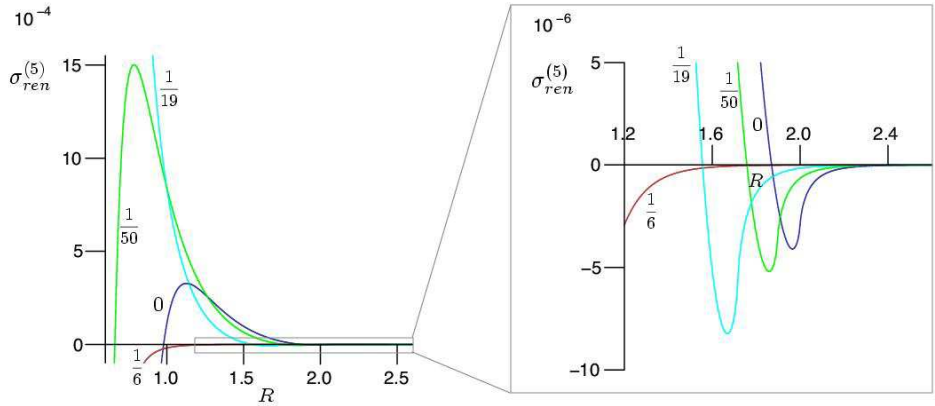,width=13cm}
\caption{$\sigma^{(5)}_{ren}$ versus $R$  for $\mu=1$ and $\xi=0, \frac{1}{50}, \frac{1}{19}, \frac{1}{6}$.}
\label{5esfera}
\end{figure}

Next we consider the case with $n=5$, for which $\rho^{(5)}_{ren}$ and $\sigma^{(5)}_{ren}$ are plotted in figures \ref{5esferarho} and \ref{5esfera}.  Again, qualitatively $\rho^{(5)}_{ren}$ and $\sigma^{(5)}_{ren}$ have similar behaviours: for sufficiently small coupling there are now two zeros. Increasing the coupling, the two zeros disappear; first the one with positive first derivative and then the one with negative first derivative. The curves then become strictly negative. Again note that the asymptotic behaviour is in agreement with  \eqref{final1gg}. To check \eqref{rhooddconf}, which in this case implies that we must have $\rho_{ren}^{(5)}<0$  for $a=0$, note that the two branches meet at $R=2\sqrt{1-5\xi}$. It is simple to check that the renormalised energy density is indeed negative for all couplings represented at the corresponding radius.

\begin{figure}[t!]
\begin{picture}(0,0)(0,0)
\end{picture}
\centering\epsfig{file=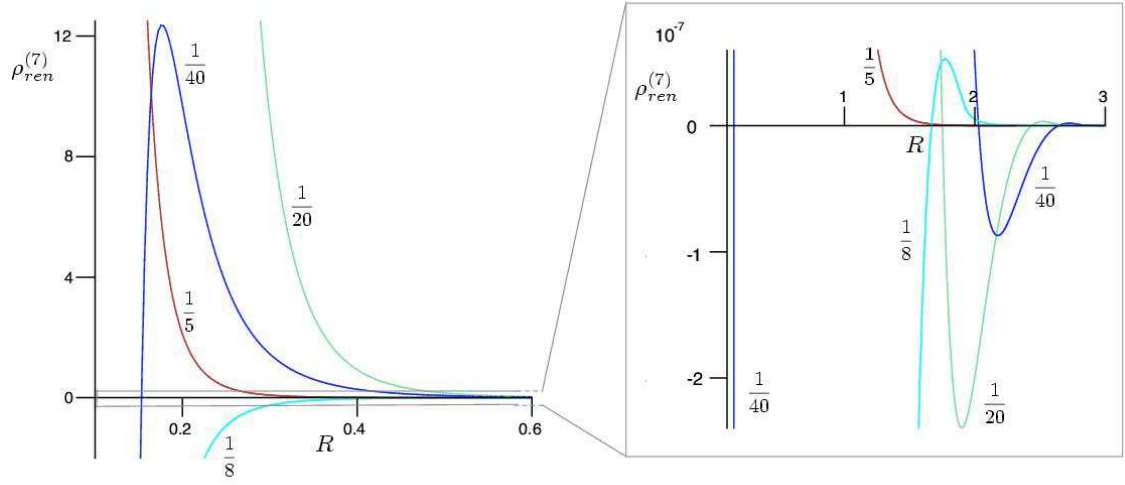,width=15cm}
\caption{$\rho^{(7)}_{ren}$ versus $R$  for $\mu=1$ and $\xi=\frac{1}{40}, \frac{1}{20}, \frac{1}{8}, \frac{1}{5}$.}
\label{7esferarho}
\end{figure}

\begin{figure}[h!]
\begin{picture}(0,0)(0,0)
\end{picture}
\centering\epsfig{file=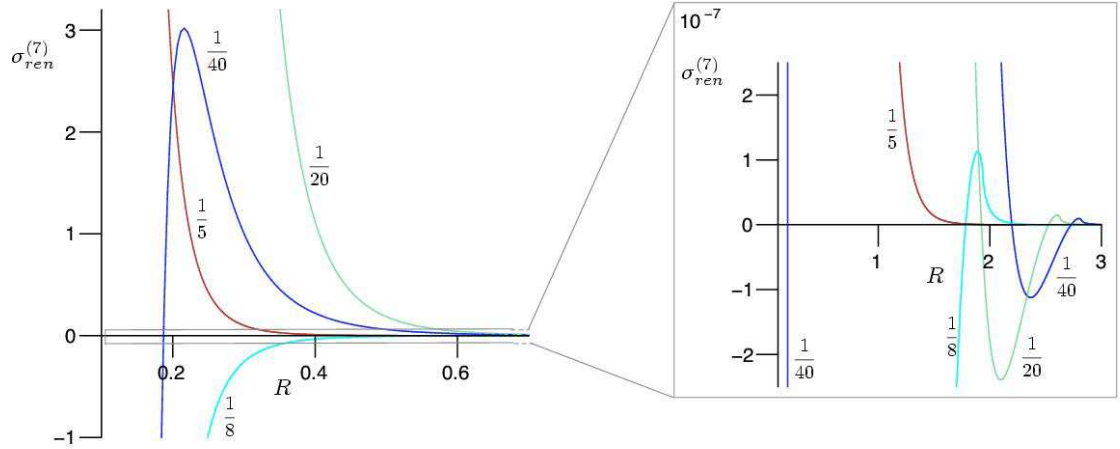,width=15cm}
\caption{$\sigma^{(7)}_{ren}$ versus $R$ for $\mu=1$ and $\xi=\frac{1}{40}, \frac{1}{20}, \frac{1}{8}, \frac{1}{5}$.}
\label{7esferasigma}
\end{figure}

Moving on to the 7-sphere, one can see a pattern starting to emerge (for odd $n$). $\rho^{(7)}_{ren}$ and $\sigma^{(7)}_{ren}$ are now plotted in figures \ref{7esferarho} and \ref{7esferasigma}. Again, the same (qualitative) behaviour is found for the two functions: one finds that, for sufficiently small coupling, there are now three zeros. As one increases the coupling, the number of zeros decreases until none remains. The asymptotic behaviour is in agreement with  \eqref{final1gg}, since $\rho^{(7)}_{ren}$ approaches zero through positive values. To check \eqref{rhooddconf}, which in this case implies that we must have $\rho_{ren}^{(7)}>0$  for $a=0$, note that the two branches meet at $R=3\sqrt{1-14\xi/3}$. It is simple to check that the renormalised energy density is indeed positive for all couplings represented at the 
 corresponding radius.

It seems tempting to generalise the observed pattern for any odd $n$ sphere. One expects, for sufficiently small coupling, that $\rho^{(n)}_{ren}$ and $\sigma^{(n)}_{ren}$ start off at negative values, possess $(n-1)/2$ zeros and vanish exponentially as $R\rightarrow +\infty$ with sign $(-1)^{(n-3)/2}$. Continuously increasing the coupling, the two quantities start to lose their zeros, one at the time and alternate the sign for small values of $R$ each time a zero is lost. For sufficiently large coupling (but always before conformal coupling) there are no zeros remaining and the sign becomes the same as the asymptotic one.

A final observation relates the behaviour of  $\rho^{(n)}_{ren}$ and $\sigma^{(n)}_{ren}$, as functions of $R$, for a given coupling in the massless and massive case: whenever the massive curve has an even  (odd) number of zeros the sign of the massless curve coincides (is the opposite) with the asymptotic sign of the massive one.

\subsection{Even $n$}
\label{evenspheres}
We have already mentioned, at the beginning of this section, that for even spheres and $a=0=\mu$ the energy density vanishes. It is interesting to start analysing the leading behaviour in $a$ for either branch of \eqref{rho_ren2} as $a\rightarrow 0$. Denoting $|a^2R^2|\equiv \epsilon^2\ll 1$ in either branch we find (up to the term $\rho_{extra}^{(2p+4)}$, which is the same in both branches)
\begin{equation}
\begin{array}{c}
\rho^{(2p+4)}_{ren}=\displaystyle{\frac{(-1)^p}{R^{2p+5}V_{R=1}^{(2p+4)}}}\left\{\begin{array}{ll}
\displaystyle{\frac{\beta_0}{6}\epsilon^3+\frac{\beta_1}{15}\epsilon^5+\mathcal{O}(\epsilon^7)-\left(\frac{\beta_0}{6}\epsilon^3-\frac{\beta_0\pi^2}{32}\epsilon^4+\frac{\beta_1}{15}\epsilon^5+\mathcal{O}(\epsilon^6)\right)}\ , &
  (a^2\ge 0) \ , \\
 \displaystyle{\frac{\beta_0\pi^2}{32}\epsilon^4 +\mathcal{O}(\epsilon^6)  \ ,}& (a^2<0) \ . 
\end{array}\right. 
\end{array} \label{rho_ren2exp}
\end{equation}
For clarity we have displayed separately the expansions of the finite sum and the integral in \eqref{rho_ren2}. Remarkably, the finite sum - which appeared naturally in our renormalisation procedure - smooths the energy density at $a=0$. This is another piece of evidence for the correctness of our procedure.

Let us start by considering the asymptotic behaviour  of the energy density. An analysis of \eqref{rho_ren2} shows that the leading term as $R\rightarrow +\infty$ is of order $1/R^{2p+6}$:
\bequ
\rho^{(2p+4)}_{ren}\stackrel{R\rightarrow+\infty}{=}\frac{(-1)^p}{4\mu R^2 V^{(2p+4)}}\sum_{m=0}^{p+1}\beta_m\left[\frac{\eta^{m+2}}{2(m+1)(m+2)}+(-1)^m\zeta_R(-2m-3)\left(1-\frac{1}{2^{2m+3}}\right)\right]+\dots \ . \label{assneven}\eequ
The second term inside the square brackets is always positive; the sign of the first term depends on $\eta$. The simplest case is conformal coupling, for which $\eta=0$; then the sign of \eqref{assneven} is $(-1)^p$. This will be readily checked in our numerical plots. For coupling smaller than the conformal value, $\eta$ becomes negative and the sign of \eqref{assneven} might differ from $(-1)^p$; indeed, we shall verify it is always negative for minimal coupling, regardless of $p$.

\begin{figure}[h!]
\begin{picture}(0,0)(0,0)
\end{picture}
\centering\epsfig{file=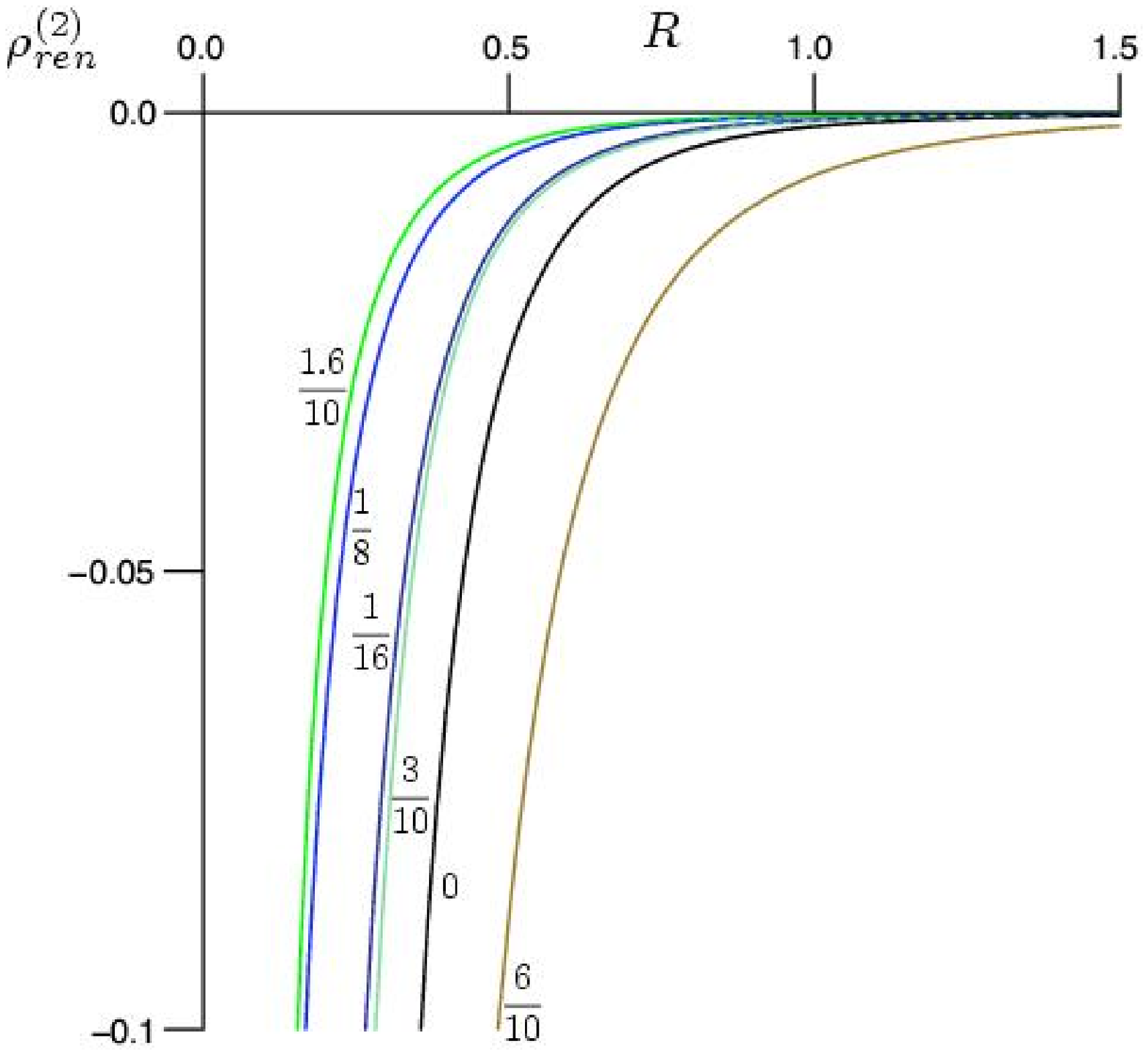,width=7.0cm} 
\centering\epsfig{file=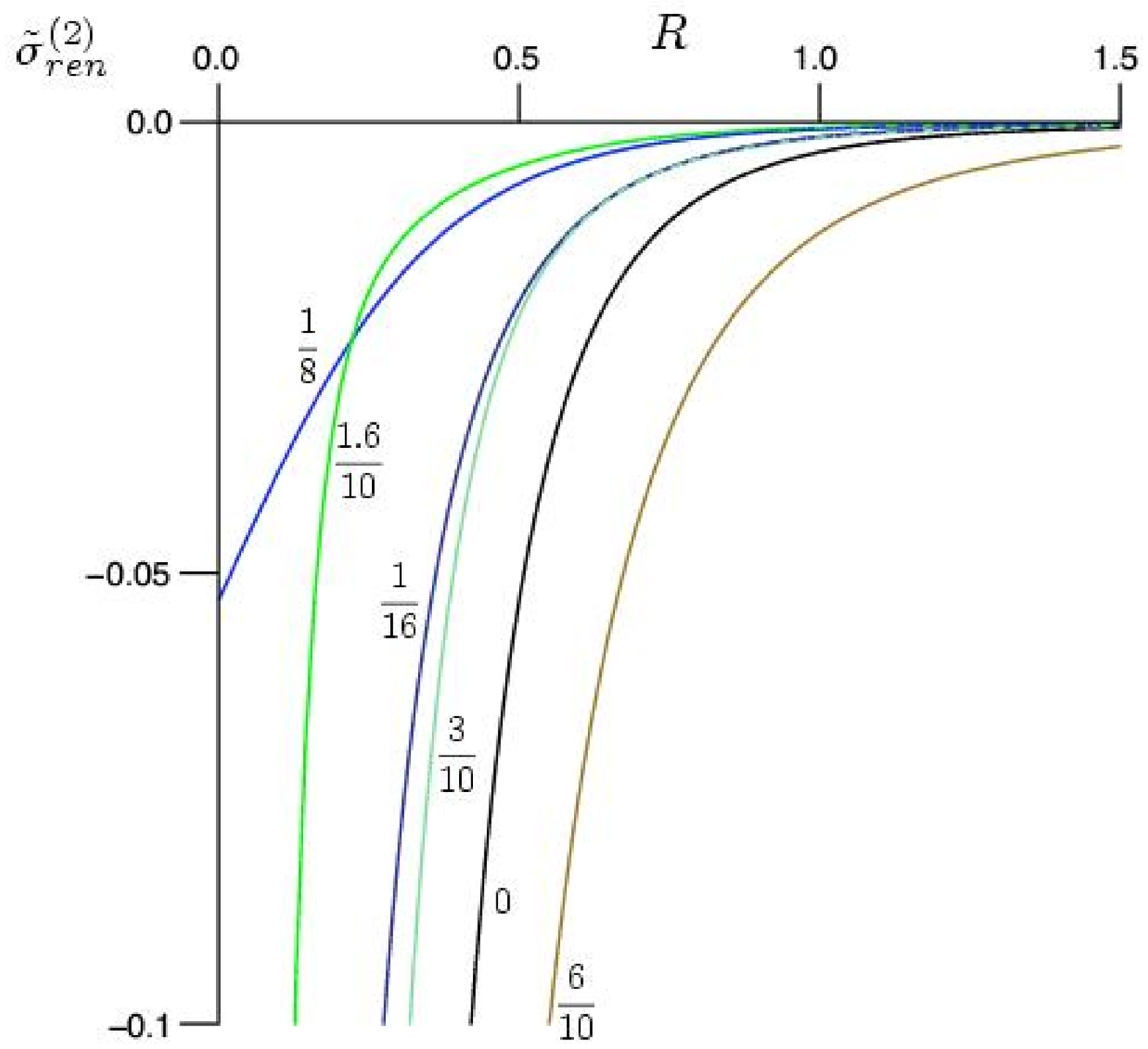,width=7.0cm}
\caption{$\rho^{(2)}_{ren}$ and $\tilde{\sigma}^{(2)}_{ren}\equiv 2p^{(2)}_{ren}$ versus radius for $\mu=1$ and $\xi=0,\frac{1}{16},\frac{1}{8},\frac{1.6}{10},\frac{3}{10},\frac{6}{10}$.}
\label{2esfera}
\end{figure}

The first non-trivial case is $n=2$, for which $\rho^{(2)}_{ren}$ and $\sigma^{(2)}_{ren}$ are displayed in figure \ref{2esfera}. A general observation is that, as for the odd $n$ case, the qualitative behaviour of $\rho^{(n)}_{ren}$ and $\sigma^{(n)}_{ren}$ is always similar. For $n=2$, the qualitative behaviour is also independent of the coupling: the curves are always negative, in agreement with \eqref{assneven}. In particular there are no zeros.

\begin{figure}[h!]
\begin{picture}(0,0)(0,0)
\end{picture}
\centering\epsfig{file=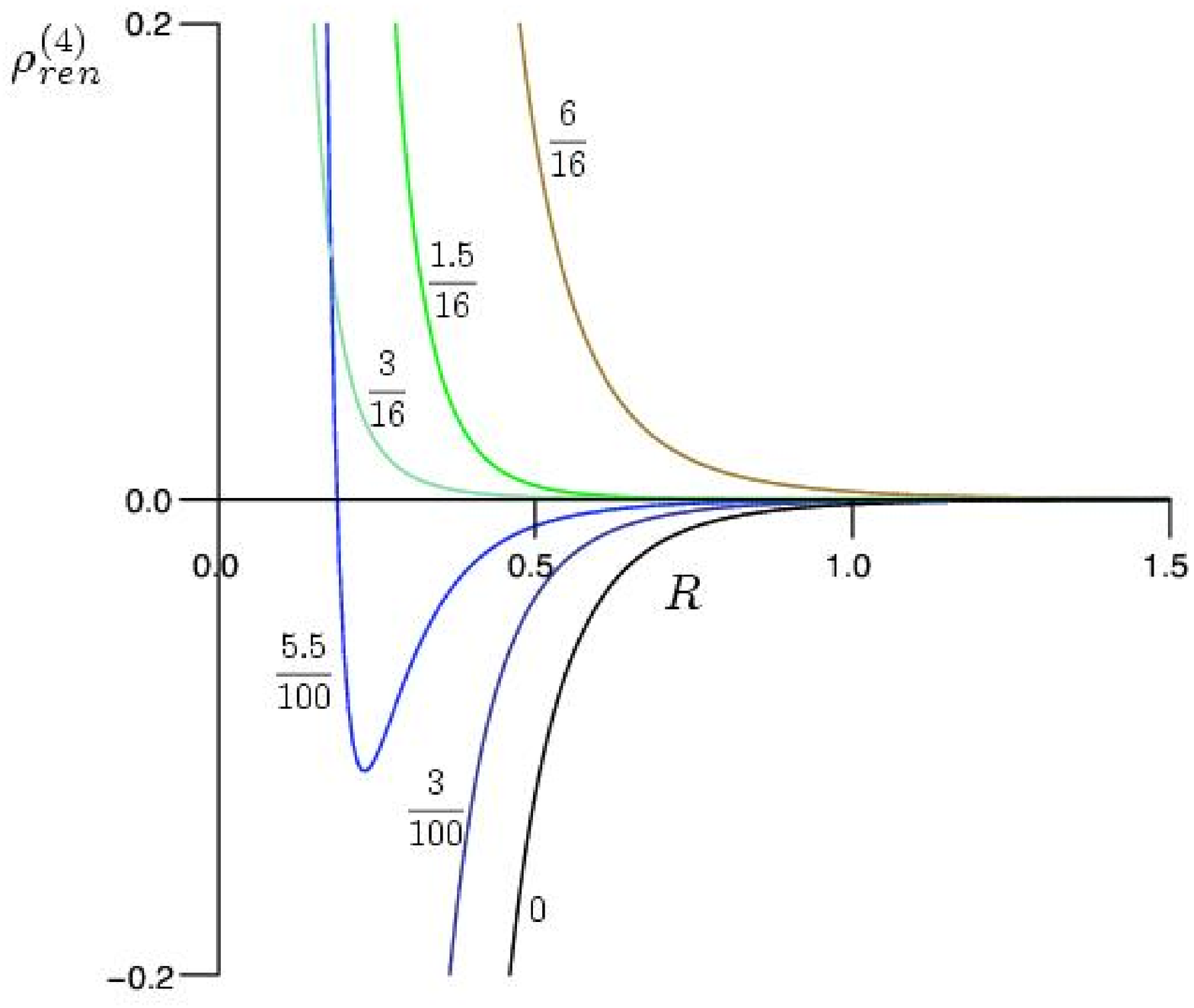,width=7.0cm} 
\centering\epsfig{file=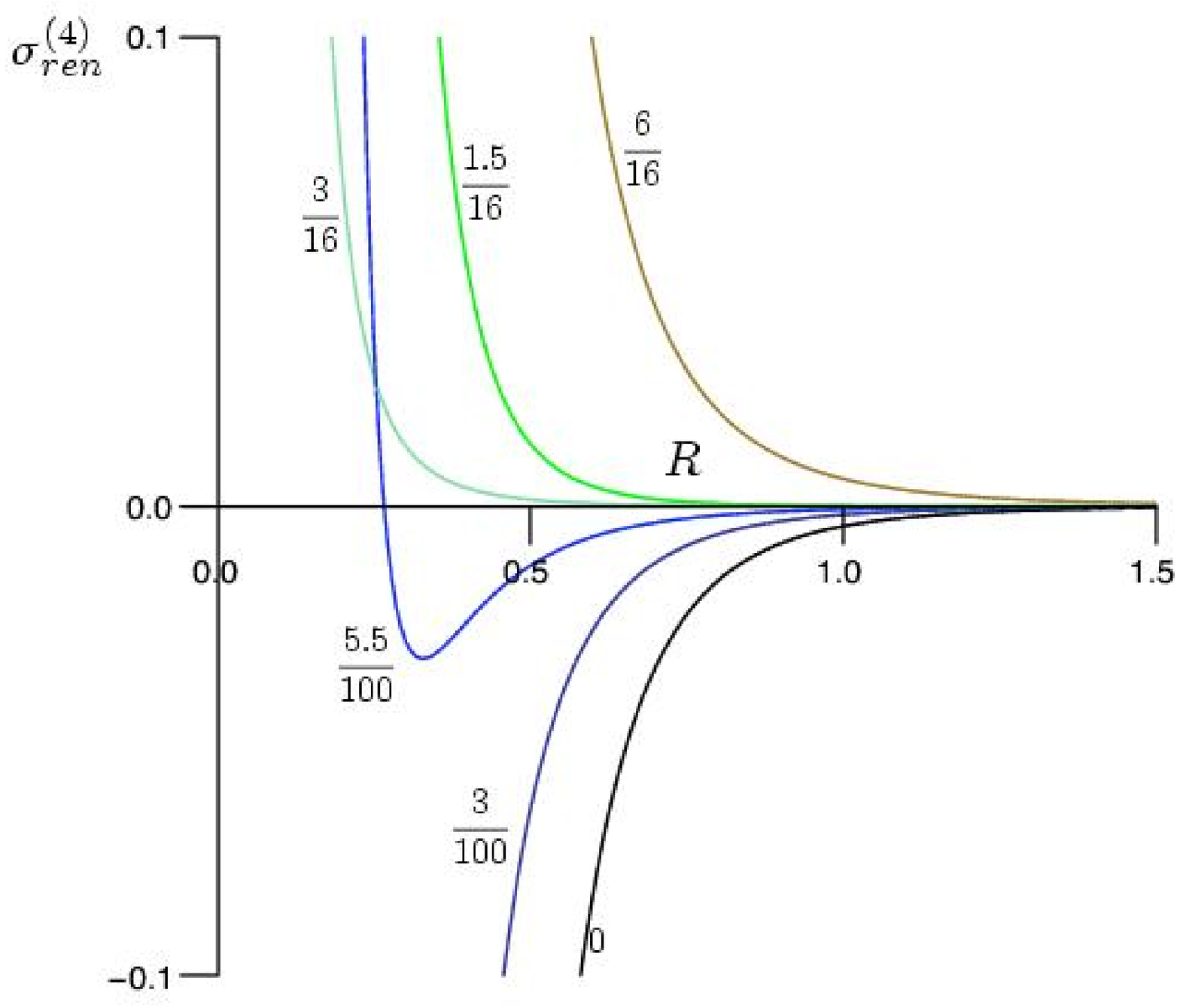,width=7.0cm} 
\caption{$\rho^{(4)}_{ren}$ and $\sigma^{(4)}_{ren}$ versus radius for $\mu=1$ and $\xi=0,\frac{3}{100},\frac{5.5}{100},\frac{1.5}{16},\frac{3}{16},\frac{6}{16}$.}
\label{4esfera}
\end{figure} 

The next case is $n=4$, for which $\rho^{(4)}_{ren}$ and $\sigma^{(4)}_{ren}$ are displayed in figure \ref{4esfera}. For minimal coupling the curves are very similar to those for $n=2$. But as one increases the coupling, the energy density and $\sigma$ become positive for sufficiently small radius, keeping their negative asymptotic value; thus, within a range of couplings these functions have one zero, which is lost beyond a certain coupling, at which the asymptotic value changes sign. Indeed, specialising \eqref{assneven} for $p=0$ we have
\[
\rho^{(4)}_{ren}\stackrel{R\rightarrow+\infty}{=}\frac{1}{96\mu R^2 V^{(4)}}\left[(8\xi-1)\left(12\xi-\frac{9}{4}\right)^2+\frac{449}{1680}\right]+\dots \ . \]
Thus, the asymptotic values changes sign for some $\xi$ smaller than the conformal value, and beyond this coupling  the curves are strictly positive.

\begin{figure}[h!]
\begin{picture}(0,0)(0,0)
\end{picture}
\centering\epsfig{file=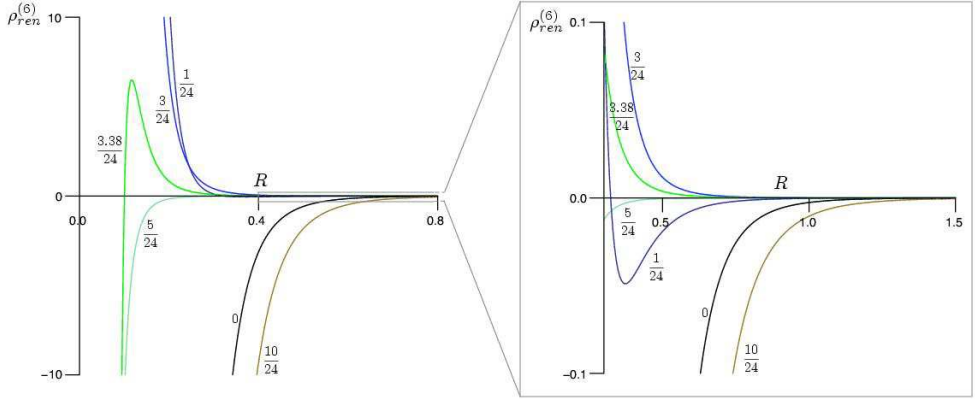,width=13.0cm}
\caption{$\rho^{(6)}_{ren}$ versus  radius for $\mu=1$ and $\xi=0,\frac{1}{24},\frac{3}{24},\frac{3.38}{24},\frac{5}{24},\frac{10}{24}$.}
\label{rho6esfera}
\end{figure} 

\begin{figure}[h!]
\begin{picture}(0,0)(0,0)
\end{picture}
\centering\epsfig{file=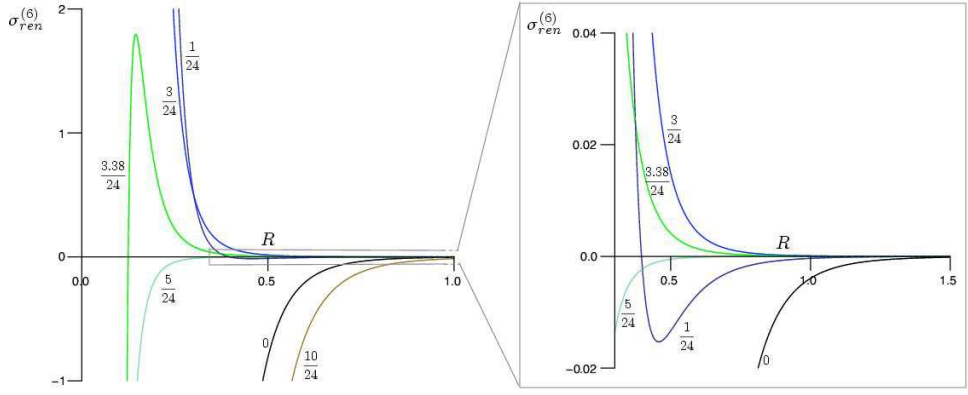,width=13.0cm}
\caption{$\sigma^{(6)}_{ren}$ versus radius for $\mu=1$ and $\xi=0,\frac{1}{24},\frac{3}{24},\frac{3.38}{24},\frac{5}{24},\frac{10}{24}$.}
\label{sigma6esfera}
\end{figure}

The case $n=6$ is considered in  figures \ref{rho6esfera} and \ref{sigma6esfera}. For zero coupling, again, the curve resembles that obtained for $n=2,4$. As one increases the coupling one observes the same behaviour discussed for $n=4$, namely that the curve becomes positive, first for small $R$ only and then for all $R$. But as one keeps increasing the coupling one now observes another inversion: the curves become again negative; first for small values of $R$ only (hence possessing a zero) and then, for sufficiently large coupling, for all $R$. This is the asymptotic behaviour expected from \eqref{assneven}.

\begin{figure}[h!]
\begin{picture}(0,0)(0,0)
\end{picture}
\centering\epsfig{file=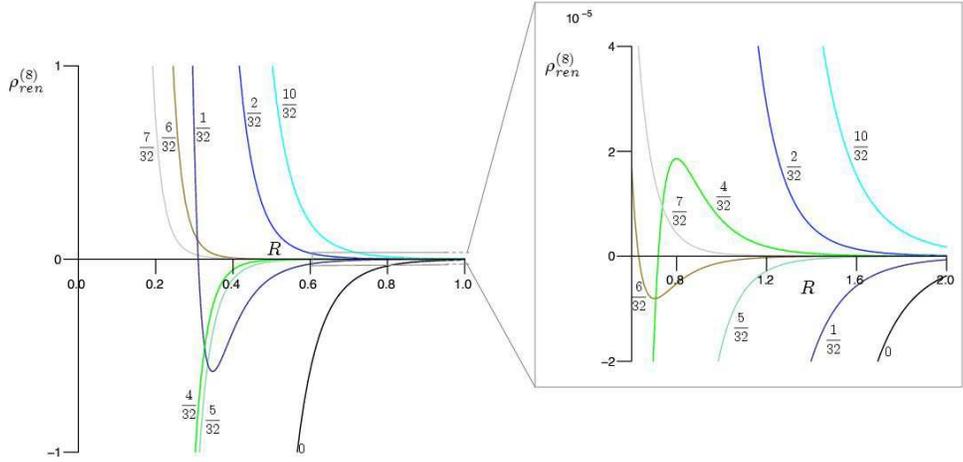,width=13.0cm} 
\caption{$\rho^{(8)}_{ren}$ versus radius for $\mu=1$ and $\xi=0,\frac{1}{32},\frac{2}{32},\frac{4}{32},\frac{5}{32},\frac{6}{32},\frac{7}{32},\frac{10}{32}$.}
\label{rho8esfera}
\end{figure} 

\begin{figure}[h!]
\begin{picture}(0,0)(0,0)
\end{picture}
\centering\epsfig{file=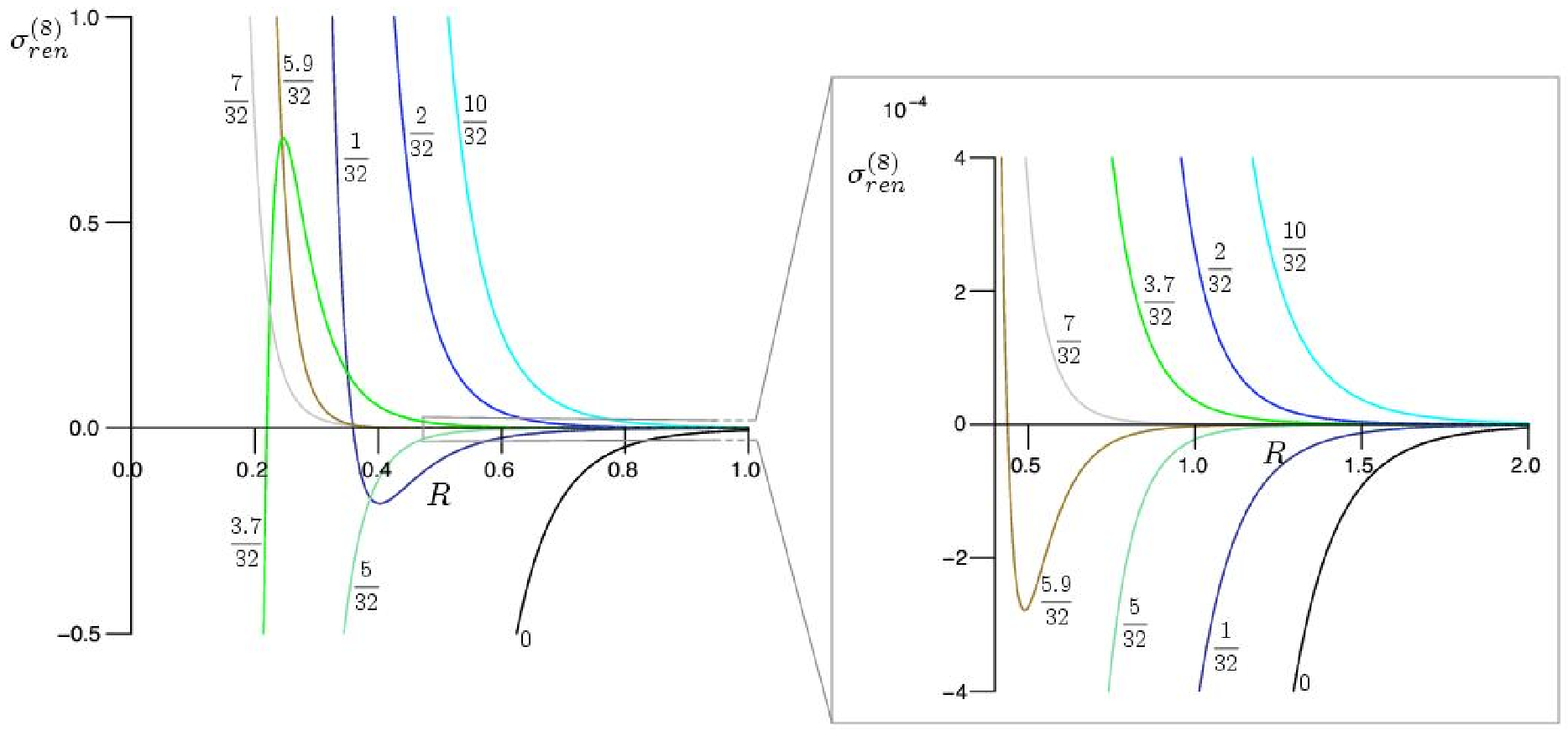,width=13.0cm} 
\caption{$\sigma^{(8)}_{ren}$ versus radius for $\mu=1$ and $\xi=0,\frac{1}{32},\frac{2}{32},\frac{3.7}{32},\frac{5}{32},\frac{5.9}{32},\frac{7}{32},\frac{10}{32}$.}
\label{sigma8esfera}
\end{figure}

The case $n=8$, which is considered in  figures \ref{rho8esfera} and \ref{sigma8esfera} confirms the emerging pattern. A further oscillation is then observed before conformal coupling, which makes the curves positive for sufficiently large coupling. Thus, for $n=2p+4$ one would expect the energy density and $\sigma$ to start off, at zero coupling, in qualitatively similar way to the cases analysed. As one increases the coupling, one will see $p+1$ changes in sign. In each case, the sign change occurs first for small radii (hence allowing the existence of one zero) and propagates for all radii as the coupling is increased. After these $p+1$ sign changes, the curves stabilise with sign $(-1)^p$ for all radii, approaching zero as $R\rightarrow +\infty$. Note that, in contrast with the odd $n$ case, each curve can have, at most, one zero.

\section{Conclusions and Discussion}
\label{dynamics}
Let us summarise the main results in this paper:
\begin{description}
\item{1 -} We have computed the renormalised energy momentum tensor
of a scalar field with mass $\mu$ and coupling $\xi$ to the scalar
curvature on an $n+1$ dimensional Einstein Static Universe with radius
$R$. The resulting renormalised energy-momentum tensor is completely specified by the renormalised energy density and pressure, which are presented in  \eqref{rho_ren1}-\eqref{p_ren2}.   In the process we establish some technical results:
\item{1.1 -} We show that different regularisation/renormalisation methods yield the same result, simply because the computation converges to a generic scheme, which effectively includes several
regularisation methods commonly used, such as damping function, point-splitting, a ``momentum'' cut-off and zeta
function (c.f. section \ref{generic_reg} and the discussion in appendix \ref{commonmethods});
\item{1.2 -} We show how, in these different renormalisation methods, various gravitational couplings (associated to gravitational operators with mass dimension up to $n+1$) are renormalised and why all the renormalised couplings should be considered in a self-consistent effective field theory treatment. 
\item{ 2 -} We unveil a rather non-trivial, but clear, pattern for the behaviour of the vacuum energy density as a function of coupling, dimension, mass and radius. We also observe a transition, in the gravitational effect caused by the vacuum energy momentum, from repulsive to attractive. These features were studied in section  \ref{analysis}, wherein a detailed numerical analysis, also matched with some analytical behaviours, was performed for the renormalised quantities $\rho^{(n)}_{ren}$ and $\sigma^{(n)}_{ren}$. These are the relevant quantities that enter the $n+1$ dimensional cosmological equations \eqref{cosmo1} and \eqref{cosmo2}. One can summarise the results of the analysis as follows:
\item{2.1 -} The generic radial dependence of $\rho^{(n)}_{ren}$ and $\sigma^{(n)}_{ren}$ is given by \eqref{newsigmarho}. This renders the mass dependence trivial and one needs only to consider one representative of the massive case (say, $\mu=1$) and the massless case. In either case a pattern emerges from the analysis of the lower $n$ cases and it is convenient to separate the analysis according to the parity of $n$.
\item{2.2 -} Massive, odd $n$: The qualitative behaviour of $\rho^{(n)}_{ren}$ and $\sigma^{(n)}_{ren}$ is similar and displayed for $n=3,5,7$ in figures \ref{3esfera}-\ref{7esferasigma}. For sufficiently small coupling, one observes that these quantities are negative for small $R$, possess $(n-1)/2$ zeros and vanish exponentially as $R\rightarrow +\infty$ with sign $(-1)^{(n-3)/2}$, in accordance with \eqref{final1gg}.  Continuously increasing the coupling, the two quantities start to drop off their zeros, one at the time, and alternate their sign for small values of $R$ each time a zero is lost. For sufficiently large coupling (but always before conformal coupling) no zeros remain and the sign becomes the same as the asymptotic one. 
\item{2.3 -} Massive, even $n$: The qualitative behaviour of $\rho^{(n)}_{ren}$ and $\sigma^{(n)}_{ren}$ is again similar, but distinct from the odd $n$ case; these quantities are displayed for $n=2,4,6,8$ in figures \ref{2esfera}-\ref{sigma8esfera}. For sufficiently small coupling these curves are negative for all $R$, vanishing as a power law, $1/R^{n+2}$, as $R\rightarrow +\infty$, in accordance with \eqref{assneven}. Continuously increasing the coupling, $n/2-1$ changes in sign occur. In each case, the sign change occurs first for small radius (hence allowing the existence of one zero) and propagates for all radii as the coupling is increased. After these $n/2-1$ sign changes, the curves stabilise with sign $(-1)^{n/2}$ for all radii, approaching zero as $R\rightarrow +\infty$. Note that, in contrast with the odd $n$ case, for each coupling there can be, at most, one zero for both $\rho^{(n)}_{ren}$ and $\sigma^{(n)}_{ren}$.   
\item{2.4 -} Massless: The radial dependence becomes the simple power law \eqref{rhomass0}. The only quantity to be determined is the coefficient $\alpha_{ren}^{(n)}$, which depends on the coupling $\xi$ and is displayed in figures \ref{alpha} for $n=3,5,7$ and figure \eqref{rho2468} for $n=2,4,6,8$. One observes that this coefficient is always negative for small $\xi$. For $n$ odd $\alpha^{(n)}_{ren}(\xi)$ has $(n-1)/2$ zeros and goes to zero for large $\xi$ with sign $(-1)^{(n+1)/2}$. For $n$ even $\alpha^{(n)}_{ren}(\xi)$ has $n/2$ zeros and diverges - as can be checked from \eqref{rho_ren2} - for large $\xi$ with sign $(-1)^{n/2}$. In this case, one of the zeros is an extreme.
\item{2.5 -} Massless and conformal coupling $\xi=\xi_{conf}$ (conformal scalar field theory): the Casimir energy density vanishes for even $n$; for odd $n$ its sign alternates as given by \eqref{rhomass0} and \eqref{rhooddconf}. This behaviour, although not the general analytic expressions, had already been noticed in \cite{ozcan}. 
\end{description}

We would like to close with a discussion of the backreaction of the vacuum energy density and stress and with the observation that a number of self-consistent ESUs can be found, in all dimensions, taking into account the quantum effects.

 Consider equations \eqref{cosmo1} and \eqref{cosmo2} with both classical and quantum terms, which are
\[
\rho^{(n)}_{cla}=\Lambda+\frac{A}{R^n} \ , \ \ \ \ \ \sigma^{(n)}_{cla}=\frac{-2\Lambda}{n-2}+\frac{A}{R^n} \ , 
\]
where $\Lambda,A>0$ represent the contribution of a positive cosmological constant and a pressure-less, positive density perfect fluid, and the quantum contribution is taken in the form \eqref{newsigmarho}. Equations \eqref{cosmo1} and \eqref{cosmo2} imply that an equilibrium radius, $R_{eq}$ will be possible iff
\bequ
\left\{ \begin{array}{l}
\displaystyle{R_{eq}^{n-1}=\frac{2[\Lambda_0+A_0 + \alpha^{(n)}_{ren}(\xi,\mu_0)]}{n(n-1)}} \vspace{0.3cm}\\
\displaystyle{\varsigma^{(n)}_{ren}(\xi,\mu_0)=\frac{2\Lambda_0}{n-2}-A_0} \ , \end{array} \right. \label{equilibrium} \eequ
where we have defined
\[
\Lambda_0=\Lambda R_{eq}^{n+1} \ , \ \ \ \ \ \ A_0\equiv AR_{eq}  \ , \ \ \ \ \ \ \mu_0\equiv \mu R_{eq} \ .  \]
To examine solutions of these equations start by considering the massless case, for which \eqref{sigmamass0} holds. Then, \eqref{equilibrium} becomes
\bequ
\left\{ \begin{array}{l}
\displaystyle{R_{eq}^{n-1}=\frac{2\left[A_0+(n+1)\Lambda_0\right]}{n(n-1)^2}} \vspace{0.3cm}\\
\displaystyle{\alpha^{(n)}_{ren}(\xi)=\frac{2\Lambda_0-(n-2)A_0}{n-1}} \ . \end{array} \right. \label{equilibriummassless} \eequ
One may think of these equations as determining, $R_{eq}$ and $\xi$ for given $A_0,\Lambda_0,n$. Generically there can be several solutions for each choice of $A_0,\Lambda_0,n$ (or none). The solution can even be the same as the classical one \eqref{classicalsol}, by fine tuning $\xi$ to be one of the zeros of $\alpha^{(n)}_{ren}(\xi)$; such zeros exist for all $n$ and, for $n$ even,  a special case is the conformal value. 

For the generic case, \eqref{equilibrium}, one may think of such equations as determining, $R_{eq}$ and $\mu_0$ for given $A_0,\Lambda_0,n,\xi$. Again, these equations can have a number of solutions, so that equilibrium radii can be found. An interesting question is if such equilibrium radii can be \textit{stable}, in view of the classical instability of the ESU against radial perturbations. The quantum contribution seems, by itself, to be able to have such effect, since the radial derivative at some of the zeros of $\sigma^{(n)}_{ren}$ is \textit{positive}, thus producing a restoring force, by virtue of \eqref{cosmo2}. But a dynamical analysis of \eqref{cosmo1} and \eqref{cosmo2} with the renormalised energy momentum computed herein must be performed to test this possibility. This is work in progress  \cite{paper2}.

\section*{Acknowledgements}
We would like to thank Filipe Paccetti for very useful discussions. M.S. is supported by the FCT grant SFRH/BD/23052/2005. R.R is supported by the FCT grant SFRH/BD/35984/2007. This work was also supported by  the FCT grant  POCTI/FNU/50161/2003. Centro de F\'\i sica do Porto is partially funded by FCT through POCTI programme.

\appendix

\section{Contact with commonly found regularisation procedures}
\label{commonmethods}
We have already seen in section \ref{generic_reg} that the general regularisation scheme used, expressed by \eqref{reged0}, makes close contact with a regularisation scheme that employs a damping function.  We now make contact with other well known methods commonly found in the literature. We will show how, using these methods, at a certain point of the computation one arrives at a special case of the general regularisation procedure considered in \eqref{reged0}. The main point is that all of them
seem to collapse into the same result, once we have introduced a
sensible cut-off scale.

\subsection{Point-splitting, an alternative formulation}
\label{pointsplit}
Point-splitting constitutes a covariant regularisation scheme, thus naturally respecting spacetime symmetries \cite{Birrel,Christensen:1976vb}. The conventional approach to this method consists in regularising the energy momentum tensor by noticing that the divergences arise from evaluating operators (which
are quadratic in the field) at the \textit{same} point. Therefore, one simply evaluates such operators at \textit{different} points. Using \eqref{EM_tensor} we can formally define,
\begin{multline}
T_{\mu
  \nu}=\lim_{x\rightarrow
  x'}\left(1-2\xi\right)\partial_{\mu}\Phi\partial_{\nu'}\Phi'+\xi\left(\Phi'\mathcal{R}_{\mu\nu}\Phi-\Phi'D_{\mu}D_{\nu}\Phi-\Phi D_{\mu'}D_{\nu'}\Phi'\right)
  \\
  +g_{\mu\nu}\left(2\xi-\frac{1}{2}\right)\left[\partial_{\alpha'}\Phi'\partial^{\alpha}\Phi+\Phi'(\mu^2+\xi\mathcal{R})\Phi\right]
  \ . \label{non_tensorial}
\end{multline}
where $x$ is replaced by $x'\neq x$ in all primed quantities. 

To make contact with the treatment in section \ref{generic_reg}, the first step is to identify the regularisation length scale $\gamma L$. This issue is intrinsically related to the issue that equation \eqref{non_tensorial} is only valid as long as we refer to a particular coordinate system and work in components; it
has \textit{no frame independent meaning} - it is only a tensor in the limit of coincidence.\footnote{In the conventional approach, one writes the point splitted energy momentum tensor in terms of Green's functions (and its derivatives) for the scalar field. The objects arising can be given a meaning as \textit{bitensors} and may be manipulated in a way to separate various contributions which transform as tensors at $x$  (they also contain the geodesic distance between $x$ and $x'$, for more details see \cite{Christensen:1976vb}).} However, as argued in section \ref{generic_reg}, we expect the fundamental theory to have an intrinsic regularisation which produces a regularised energy momentum \textit{tensor}. 

To construct a true point splitted regularised tensor we proceed as follows. We start by considering $x'$ as a fixed map, which plays the role of a regularisation parameter. To avoid confusion we define such a map as
\begin{eqnarray}
h:& \mathcal{M}&\rightarrow \mathcal{M}  \nonumber\\
& x&\rightarrow h(x) \ , \nonumber
\end{eqnarray}
where $\mathcal{M}$ is the background manifold and $x$ is a point on
the manifold.
This map induces a pullback map, which associates a $\binom{0}{q}$ tensor at
$x$ to every $\binom{0}{q}$ tensor at $h(x)$. By taking the map $h$ to be the identity we recover the original
tensor field. Therefore this is a very natural way of constructing a
point splitted field. 

For a scalar field, this procedure is particularly simple since the
pullback is given by composition  of the scalar field with the
map $h$,
\begin{eqnarray}
\Phi_h:& \mathcal{M}&\rightarrow \mathbb{R} \nonumber \\
& x&\rightarrow \Phi_h(x)\equiv\Phi(h(x)) \ . \nonumber
\end{eqnarray}
So if we go back to \eqref{EM_tensor} we can obtain a true point splitted
energy momentum tensor by replacing one of the $\Phi(x)$'s by
$\Phi_h(x)$ in each quadratic term in $\Phi$. Similarly we
could have started by performing the substitution at the level of the
Lagrangian and consider it the regularised Lagrangian for the
theory. Functional differentiation would then yield the energy
momentum tensor. In any case the result is exactly the same
\begin{multline}
T_{\mu
  \nu}^{\Phi,h}=(1-2\xi)\partial_{\mu}\Phi\partial_{\nu}\Phi_h+\xi\left(\Phi_h\mathcal{R}_{\mu\nu}\Phi-\Phi_h
D_{\mu}D_{\nu}\Phi-\Phi
D_{\mu}D_{\nu}\Phi_h\right)+\\+g_{\mu\nu}\left(2\xi-\frac{1}{2}\right)\left[\partial_{\alpha}\Phi_h\partial^{\alpha}\Phi+\Phi_h(\mu^2+\xi\mathcal{R})\Phi\right] \ , \label{ttensor}
\end{multline}
with the  bare energy momentum tensor being recovered by taking
\[
T_{\mu \nu}^{\Phi}=\lim_{h\rightarrow i_d}T_{\mu\nu}^{\Phi,h} \ , 
\]
where $i_d$ is the identity map.

Expression \eqref{ttensor} is a manifestly covariant quantity and can be put in a form which makes contact with the conventional point splitting technique,
if we express the covariant derivative terms in new coordinates
$x'=h(x)$, i.e.
\[
D_\mu\equiv \dfrac{dh^{\alpha'}(x)}{dx^\mu}D_{\alpha'} \ , 
\]
we obtain an expression similar to \eqref{non_tensorial}
\begin{multline}
T_{\mu\nu}^{\Phi,h}=(1-2\xi)\dfrac{dh^{\alpha'(x)}}{dx^{\nu}}\partial_{\mu}\Phi\partial_{\alpha'}\Phi'+\xi\left(\Phi'\mathcal{R}_{\mu\nu}\Phi-\Phi'
D_{\mu}D_{\nu}\Phi-\dfrac{dh^{\alpha'(x)}}{dx{^\mu}}\dfrac{dh^{\beta'(x)}}{dx^{\nu}}\Phi
D_{\alpha'}D_{\beta'}\Phi'\right)+\\+g_{\mu\nu}\left(2\xi-\frac{1}{2}\right)\left[\dfrac{dh^{\beta'(x)}}{dx^{\alpha}}\partial_{\beta'}\Phi'\partial^{\alpha}\Phi+\Phi'(\mu^2+\xi\mathcal{R})\Phi\right] \label{point_spliting_me}\ , 
\end{multline}
where $\Phi'\equiv \Phi(x')=\Phi_h(x)$. The extra Jacobian factors
make this quantity a true tensor; so in fact, we have shown that the point-splitting technique can be
put in a truly frame independent way. The conventional point
splitted energy momentum tensor \eqref{non_tensorial} is obtained if 
\[
\dfrac{dh^{\alpha'(x)}}{dx^{\mu}}=\delta^{\alpha'}_{\mu} \ ;
\]
thus,  the conventional procedure just amounts to take the map $h$ to be an (arbitrary) constant translation. The regularisation length scale $\gamma L$ can now be seen as the proper size of this constant translation vector. 

We can proceed to compute the vev of $T_{\mu\nu}$ by applying the standard procedure, as described in \cite{Birrel}, and make contact with the generic regularisation in section \ref{generic_reg}. The usual method consists of  expressing such a vev in terms of the Hadamard elementary function. Here, however, we prefer to use an expansion in mode functions together with the action of the differential operators on the latter. For concreteness we focus our attention on the 3-sphere, for we can simplify the treatment using group theory. Write the $S^3$ metric as
\[
  ds_{S^3}^2=\dfrac{R^2}{4}\left[(\sigma_L^1)^2+(\sigma_L^2)^2+(\sigma_L^3)^2\right]=\dfrac{R^2}{4}\left[(\sigma_R^1)^2+(\sigma_R^2)^2+(\sigma_R^3)^2\right] \ ,\]
where $\sigma_L^i$ ($\sigma_R^i$) are left (right) forms on $SU(2)\simeq S^3$.\footnote{To see an explicit representation in terms of Euler angles see \cite{Gauntlett:2002nw}, appendix A. In the notation therein, the complexified Killing vectors used herein are ${\bf k}_j^L=-i\xi^L_j$ and ${\bf k}_j^R=i\xi^R_j$.}  The isometry group is six dimensional; the Killing vectors form the Lie algebra  of $SO(4)$, which decomposes into $\mathfrak{su}(2)_L\times
 \mathfrak{su}(2)_R$. Thus we can always choose mode functions transforming under irreducible representations of the algebra.
 
For the 4-dimensional ESU, consider the (complexified) Killing
 vectors $\{\bold{k}_0,\bold{k}_i^L, \bold{k}_i^R\}$ where $\bold{k}_0\equiv -i\partial_t$ is the timelike Killing vector and the remaining ${\bf k}$'s are the dual vector fields to the $\sigma$'s. The full isometry group of the ESU has Lie algebra  $\mathfrak{L}\equiv\mathfrak{u}(1)\times\mathfrak{su}(2)_L\times\mathfrak{su}(2)_R$. Thus, functions which transform as irreducible
representations of $\mathfrak{L}$ can be chosen as mode functions. Note also that the Casimir operator $\bold{k}^2$ of each of the $\mathfrak{su}(2)$ algebras is in fact the same, so the action on a generic basis function
$\left|\omega,\ell,m_L,m_R\right>\equiv\left|\omega\right>\otimes\left|\ell,m_L,m_R\right>$
is \cite{edmonds}
\begin{eqnarray}
\bold{k}_0\left|\omega,\ell,m_L,m_R\right>&=&\omega\left|\omega,\ell,m_L,m_R\right> \ , 
\nonumber\\
\bold{k}^{L}_1\left|\omega,\ell,m_L,m_R\right>&=&C_+(\ell,m_{L})\left|\omega,\ell,m_L+1,m_R\right>+C_-(\ell,m_{L})\left|\omega,\ell,m_L-1,m_R\right> \ , \nonumber
\\
\bold{k}^{L}_2\left|\omega,\ell,m_L,m_R\right>&=&-iC_+(\ell,m_{L})\left|\omega,\ell,m_L+1,m_R\right>+iC_-(\ell,m_{L})\left|\omega,\ell,m_L-1,m_R\right>\ , \nonumber\\
\bold{k}^{L}_3\left|\omega,\ell,m_L,m_R\right>&=&m_{L}\left|\omega,\ell,m_L,m_R\right>\ , \nonumber
\\
\bold{k}^{2}\left|\omega,\ell,m_L,m_R\right>&=&\ell(\ell+1)\left|\omega,\ell,m_L,m_R\right>\ , \label{killing_action}
\end{eqnarray}
with similar expression for the $R$ algebra (interchanging $L$ and $R$) and where 
\[
C_\pm(\ell,m)\equiv \sqrt{\ell(\ell+1)-m(m\pm1)} \ . 
\]
 The quantities $\{\omega\in \mathbb{R},\ell\in \mathbb{N}_0\}$ classify the sub-algebras and are interpreted
 respectively as energy and orbital angular
 momentum eigenvalues and
 $m_{R,L}\in \{-\ell,-\ell+1, ...,\ell\}$ are the respective azimuthal quantum numbers. Note
 that $\left|\omega\right>=e^{-i\omega t}$ and that the spatial
 functions $Y_\alpha(\underline{x})\equiv \left|\ell,m_L,m_R\right>$ obey the same differential equation
 as the hyper-spherical harmonics. The index $\alpha$
 denotes the set $\{\ell,m_L,m_R\}$ and $\underline{x}$ is a unit vector
 in the embedding space of $S^3$ ($\mathbb{R}^4$), defined by
 the angles that we choose to parameterise the sphere. 

By taking into account the dispersion relation, we finally recover the
original eigenfunctions of the Klein-Gordon equation
\begin{equation}
\phi_\alpha(t,\underline{x})=\left|\omega_\ell,\ell,m_L,m_R\right>=\dfrac{1}{\sqrt{2\omega_\ell}}e^{-i\omega_\ell
t}Y_\alpha(\underline{x})\label{killing_eigen} \ , 
\end{equation}
where we have assumed normalisation of $Y_\alpha(\underline{x})$ on the
3-sphere,
\[
\int{d^3\underline{x}\,Y_\alpha(\underline{x})Y^
*_{\alpha'}(\underline{x})}=\delta_{\alpha\alpha'} \ , 
\]
and of $\phi_\alpha(t,\underline{x})$ with respect to the Klein-Gordon
scalar product,
\[
\left(\phi_\alpha,\phi_{\alpha'}\right)\equiv
\int{d^3\underline{x}\,\left\{\phi_\alpha(t,\underline{x})\bold{k}_0\phi^*_{\alpha'}(t,\underline{x})-\phi^*_{\alpha'}(t,\underline{x})\bold{k}_0\phi_\alpha(t,\underline{x})\right\}=\delta_{\alpha\alpha'}}\ .
\]
The advantage of choosing a non-coordinate basis of Killing vectors is that all the partial derivatives are replaced by Killing vectors which have a simple action on the basis functions. The point
splitted energy density then takes the form
\begin{multline}
 2T_{ab}^{\Phi,h}=-(1-2\xi)\bold{k}_{a}\Phi\bold{k}_{b'}\Phi'+\xi\left(\mathcal{R}_{ab}\Phi\Phi'+\Phi\bold{k}_{a'}\bold{k}_{b'}\Phi'\right)+\\
g_{ab}\left(2\xi-\frac{1}{2}\right)\left[-g^{ec}\delta^{d'}_{e}\bold{k}_c\Phi\bold{k}_{d'}\Phi'+(\mu^2+\xi\mathcal{R})\Phi\Phi'\right]
+x\leftrightarrow x' \ , 
\end{multline}
where ${\bf k}_a$ denotes the set $\{{\bf k}_0,{\bf k}_i^{L}\}$ or $\{{\bf k}_0,{\bf k}_i^{R}\}$ and $x\leftrightarrow x'$ denotes symmetrisation with respect to
interchange of primed with unprimed variables (including
indices). Note that we have denoted some of the $a$ and $b$ indices in primed
form to avoid the use of Kronecker deltas which would make the notation
denser. Finally, after some algebra, the vev of the energy
density in the vacuum follows from using \eqref{killing_action} 
\[
\rho_{h}=\frac{1}{2}\sum_\alpha
\omega_\ell^2\left\{\phi_\alpha(t,\underline{x})\phi^*_\alpha(t',\underline{x'})+x\leftrightarrow
x'\right\}\ .
\]
This result is very simple and suggestive. It is simply a sum over all
modes of the frequency squared times the (symmetrised) point splitted
amplitude for the respective mode. Note that, since this statement does not depend
on the number of dimensions or the basis of functions, it should be easily generalisable for all spheres. Note that \eqref{killing_eigen} also holds for any
sphere, so  in addition the sum over $\ell$ can be factored out to get
\begin{equation}
\rho_h^{(n)}=\frac{1}{2V^{(n)}}\sum_{\ell=0}^{+\infty}d^{(n)}_{\ell}\omega^{(n)}_{\ell}g_\ell(x,x') \label{reg_point_split}\ ,
\end{equation} 
where
\[
g_\ell(x,x')\equiv\cos\left(\omega^{(n)}_\ell(t-t')\right) {}_2F_1\left[-\ell,\ell+n-1,\ell+\frac{1}{2};\frac{1-\underline{x}.\underline{x}'}{2}\right] \ ;
\]
we have used the generalised addition theorem for spherical
harmonics \cite{hyper_harmonics}
\[
\sum_{\alpha\backslash\{\ell\}}Y_\alpha(\underline{x})Y^*_\alpha(\underline{x}')=\dfrac{2\ell+n-1}{(n-1)V^{(n)}}C_\ell^{\frac{n-1}{2}}(\underline{x}.\underline{x}') \ , 
\]
and a relation between the Gegenbauer functions
($C^\lambda_\nu$) and the hyper-geometric function ${}_2F_1$ \cite{gegenbauer}
\[
C^\lambda_\nu(z)=\dfrac{\Gamma(\nu+2\lambda)}{\Gamma(2\lambda)\Gamma(\nu+1)}{}_2F_1\left(-\nu,\nu+2\lambda,\lambda+\frac{1}{2};\frac{1-z}{2}\right) \ , 
\]
where the only restriction is that $\lambda$ is not a negative
semi-integer. Equation \eqref{reg_point_split} says that the regularised energy density falls into 
a similar type to the generic regularisation considered in the text \eqref{reged0}
(note that $g\rightarrow 1$ when $x\rightarrow x'$ ). The Abel-Plana
formula can too in this case be applied. The quantity
obtained remains finite when
$\underline{x}=\underline{x}'$ which shows that a
timelike point separation still regularises the result. Since any
finite contribution to the final result cannot depend on the map $h$,
we are free to make such a choice. Thus, finally evaluating ${}_2F_1$ at $\underline{x}=\underline{x}'$, which is 1, we obtain that
 \eqref{reg_point_split} is exactly of the type of  \eqref{reged0}, with
\[
g(\gamma L \omega_{\ell}^{(n)})=\cos\left(\gamma L\omega_\ell^{(n)}\right) \ , \]
and therefore we have shown how the point splitting technique overlaps with our generic class of regularisations. All the results in section \ref{section_ren_T} follow straightforwardly.

\subsection{``Momentum'' cut-off and a class of damping functions}
\label{mcutoff}
A more explicit regularisation scheme would be to apply a frequency
cutoff, which is in the same spirit as the standard momentum cut-off
used in QFT. It would consist on stopping the sum at a given $\ell=N$
which is then taken to infinity. This still contains the physical
condition that high frequencies are suppressed, even though such a
rigid regularisation seems unlikely to occur from the full
theory. The cutoff in $N$ induces a cutoff energy
$\omega^{(n)}_N$ which we identify with the inverse scale $\gamma L$ and can be implemented by introducing a step function, $\theta(z)$, in \eqref{reged0}
\[ g(\gamma L \omega_{\ell}^{(n)})=\theta(1-\omega^{(n)}_\ell/\omega^{(n)}_N) \ . \]
The application of the Abel-Plana formula require using an analytic continuation of the function $g$; thus we should take the Heaviside step function as a limit of a family of smooth functions. Taking
\begin{equation}
g(\gamma L \omega_{\ell}^{(n)})=\dfrac{1}{1+e^{-(1-\omega^{(n)}_\ell/\omega^{(n)}_N)/t}} \ , 
\end{equation}
for fixed $t$, we have a family of damping functions. The rigid momentum cut-off is obtained in the limit $t\rightarrow 0$. The results in section 2 now follow for all $t$.

Alternatively, we could use the following generalisation of the Abel-Plana formula (see e.g. \cite{Saharian:2007ph})
\begin{multline}
\sum_{m=0}^{N} G(m)=\int_0^{N}G(t)dt+\frac{G(N)}{2}+ \\
+\frac{G(0)}{2}+i\int_0^{+\infty}\frac{G(it)-G(-it)+G(N+it)-G(N-it)}{e^{2\pi t}-1}dt \ . \label{cutoffN}
\end{multline}
In principle we could do similar manipulations as before and obtain
analogous identifications for the divergences when applying
\eqref{cutoffN} to \eqref{bare_energy_density} with the sum up to
$N$.

\subsection{Zeta function}\label{apzeta}
Zeta function regularisation of \eqref{bare_energy_density} is similar in spirit to dimensional
regularisation, since it relies on analytic continuation. It is implemented by writing
\bequ
\rho_s^{(n)}= \frac{1}{2V^{(n)}}\sum_{\ell=0}^{+\infty}\frac{d_{\ell}^{(n)}}{\left(\omega_{\ell}^{(n)}\right) ^{s}}\ \nu^{s+1}= \sum_{i}c_i\zeta(f_i(s)) \ ,
\label{zeta1} \eequ
where $\zeta$ represents the appropriate zeta function describing the spectrum and $c_i$ some constant factors. Commonly found cases are the Riemann zeta function, the Hurwitz zeta function and the Epstein-Hurwitz zeta function. The renormalised energy density is then obtained taking the limit
\[ \rho_{ren}^{(n)}=\lim_{s\rightarrow -1}\rho_s^{(n)} \ , \]
and using the analytic continuation of the appropriate zeta
function. In many cases the result is immediately finite; the analytic
continuation of the zeta function ``swallows'' the divergences. But it
might happen that the analytic continuation of the zeta function has
poles at $s=-1$, in which case one needs to discuss the surviving
divergences. As an example we consider below $n$ odd and generalise the method applied in \cite{Herdeiro:2005zj}, which uses the  Epstein-Hurwitz zeta function,  to treat the case with $a^2>0$.

The regularised expression \eqref{zeta1} can, alternatively, be treated as a special case of the general treatment in section 2.4. To do that with the single parameter $\gamma$ used therein we relate the parameters $\nu$ and $s$ by taking $s+1=\omega_{\ell}^{(n)}/\nu$.  Then, we note that \eqref{zeta1} can be written in the form \eqref{reged0}, with 
\[
g(\gamma
L\omega_{\ell}^{(n)})=\left(\frac{\omega_{\ell}^{(n)}}{\nu}\right)^{-\omega_{\ell}^{(n)}/\nu}
\ , \]
where $\gamma L$ is identified with $\nu^{-1}$. Thus removing the regulator $s\rightarrow -1$ corresponds to $\nu\rightarrow +\infty$ and thus to $\gamma\rightarrow 0$, as expected. The function $g$ becomes a damping function to which the treatment in section 2 can be readily applied; the results therein follow.

\subsubsection{Standard analytic continuation}
The typical procedure in zeta function or dimensional) regularisation is analytical continuation. The latter often hides the divergences and their physical interpretation as we have emphasised. For completeness we illustrate how, for $n$ odd and $a^2>0$ we can  perform the manipulations which lead to
the analytical continuation and check that we recover the same result, albeit represented in a
different way. 

Using \eqref{frespe2} and \eqref{frespe21}, \eqref{zeta1} becomes, after a simple manipulation, 
\[
\displaystyle{\rho_{s}^{(2p+3)}=\dfrac{\mu^{s+1}}{R^{-s}V^{(2p+3)}}\sum_{\ell=0}^{+\infty}{\dfrac{1}{\left(2p+2\right)!} \prod_{j=0}^{p}{\dfrac{\left[ \left(\ell+p+1\right)^2-j^2  \right]}{\left[\left( \ell+p+1  \right)^2+\left(aR\right)^2     \right]^{s/2}}}}       }\ \ ,
\]
which can be written  as a linear combination of Epstein-Hurwitz zeta functions
\bequ
\zeta_{EH}(s,b^2)\equiv \sum_{n=1}^{+\infty}\frac{1}{(n^2+b^2)^s} \ , \eequ
plus a finite sum
\begin{equation}
\begin{array}{ll}
\rho_{s}^{(2p+3)}=\dfrac{\mu^{s+1}}{R^{-s}V^{(2p+3)}\left(2p+2\right)!} &\left\{\displaystyle{\sum_{m=0}^{p+1}{{c}_{m}\,\zeta_{EH}{\left(\frac{s}{2}+m-p-1,\left(aR\right)^2\right)}}}+\right. \\
\\
&\left.\displaystyle{-\sum_{n=1}^{p}{\left[\sum_{m=0}^{p+1}{{c}_{m}\, \left(n^2+\left(aR\right)^2\right)^{p+1-\frac{s}{2}-m}}\right]}         }\right\}\ \ ,
\label{eq:zetaehgeral}
\end{array}
\end{equation}
where the coefficients ${c}_m$ are given by 
\[
{c}_{0} =1 \ , \ \ \ \ {c}_{m}  =(-1)^{m}\sum_{ \stackrel{r_{1}, r_{2}, \cdots, r_{m}=0}{r_{1}<r_{2}<\cdots<r_{m}}}^{p}\left[\left(  r_{1}^2+\left(aR\right)^2\right)  \cdots \ \left(r_{m}^2+\left(aR\right)^2\right)\right]  \ , \  m\in \mathbb{N} \ .
\]
One can show that the last term in \eqref{eq:zetaehgeral} vanishes, in the limit $s\rightarrow -1$. As long as $a^2>0$ we may use the analytic continuation of the Epstein-Hurwitz zeta function (see e.g. \cite{nesterenko}), obtaining the result 
\bequ
\begin{array}{ll}
  \rho_{ren}^{(n)}=& \dfrac{1}{RV^{(n)}\left(n-1\right)!} \left\{\displaystyle{\sum_{m=0}^{\frac{n-1}{2}}}{c}_{m}\left[ \dfrac{2\pi^{\frac{2m-n}{2}}(aR)^{\frac{n+1-2m}{2}}}{\Gamma(\frac{2m-n}{2})}   \displaystyle{  \sum_{n=1}^{+\infty}{n^{\frac{2m-n-1}{2} }K_{\frac{2m-n-1}{2} }\left( 2\pi n aR\right)   }}\ \ \right.\right. \\
\\
&  \left.\left. \ \displaystyle{~~~~~~~~~~~~~~~~~~~~~~~~~~~~~~~~~ -\dfrac{(aR)^{n-2m}}{2} + \frac{\sqrt{\pi}}{2}\ \dfrac{\Gamma\left(\frac{2m-n-1}{2}  \right)}{\Gamma(\frac{2m-n}{2})} (aR)^{-2m+n+1}}\     \right]\right\}\  ,
\end{array}
\label{eq:zetaehgeralfinal}
\eequ
for $n\geqslant3$ and $a^2>0$, where $K_{\alpha}$ are modified Bessel functions of $\alpha^{th}$ order. The last term in \eqref{eq:zetaehgeralfinal} is divergent, due to the poles of the gamma function. It should be dropped out and understood as a renormalisation of the gravitational couplings of all the terms in \eqref{action3}, except the cosmological constant. This illustrates how different renormalisation methods give rise to different renormalisation of physical quantities, since with a damping function method of the type used in \cite{Ford2,Herdeiro:2005zj} only the cosmological constant gets renormalised (a result that generalises for arbitrary odd $n$). This is a consequence of the different UV completions of the theory obtained from different regularisation methods. But for a renormalisable theory the finite contribution should be independent of the regularisation method, which is what we are obtaining here.

Subtracting these infinite contributions one obtains the renormalised energy density. In the special case of the 3-sphere it reduces to the result obtained in \cite{Herdeiro:2005zj}, where it was shown that it coincides with the result obtained with a damping function method.  That \eqref{eq:zetaehgeralfinal} yields, in general, the same result as the one obtained before with our more generic regularisation scheme can now be checked numerically, case by case, which we have done for the 5-sphere and 7-sphere.

Finally let us mention that, for $\mu=0$ and conformal coupling, we have checked that the analytic continuation of the Hurwitz zeta function yields \eqref{rhooddconf}.

\section{Simplifying the integral I}
\label{Iap}
In this appendix we simplify the integral 
\[
I\equiv i\int_0^{+\infty}\frac{f_\gamma(it+\delta)-f_\gamma(-it+\delta)}{e^{2\pi t}-1}dt \ , 
\]
which appears in \eqref{reg_abelplana}, leading to the expressions in \eqref{int1} and \eqref{int2}. 
\begin{figure}
\begin{picture}(0,0)(0,0)
\end{picture}
\centering\epsfig{file=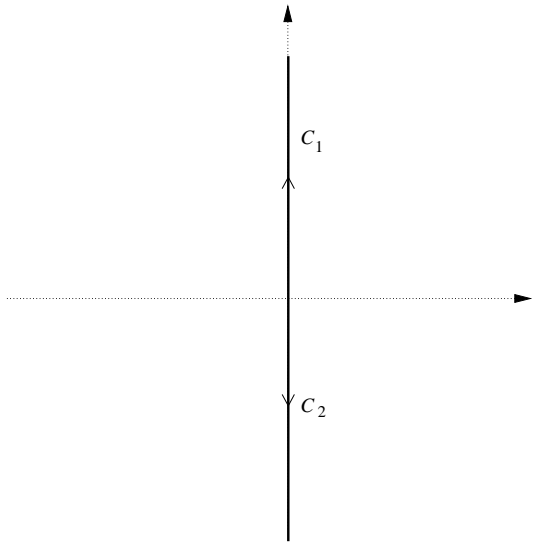,width=7cm}
\caption{Integration contours in  \eqref{I}. $C_1/C_2$ is along the positive/negative imaginary axis.}
\label{path0}
\end{figure} 
First consider the complex variable $z=it$. The integral $I$ becomes the contour integral
\bequ
I=\int_{C_1}dz\dfrac{f_\gamma\left(\delta+z\right)}{\exp(-2\pi iz)-1}+\int_{C_2}dz\dfrac{f_\gamma\left(\delta+z\right)}{\exp(2\pi iz)-1} \ , \label{I} \eequ
where the paths of integration are exhibited in figure \ref{path0}. We will deform these contours of integration into closed contours enclosing regions wherein the integrands are analytic functions. An application of Cauchy's theorem then yields the integral $I$ in terms of the integrals along the remaining paths of the closed contours providing a manifestly real expression for $I$. To do this rewrite $f_\gamma$ given in \eqref{reged} as
\[
f_\gamma(t)=\frac{1}{2}P_-(t)\sqrt{t^2+a^2R^2}g\left(\frac{\gamma L}{R}\sqrt{t^2+a^2R^2}\right) \ , 
\]
where $P_-(t)$ is given by (\ref{Pt}). It is
  convenient to analyse separately the cases with $a^2$ non-negative
  and negative respectively. Note that we will implicitly use
  the fact that the regulator is analytical. For convenience, we will keep
  $\gamma\neq 0$ for some integrals; this is allowed for all the extra contributions are of order $\gamma^k$ with $k$ positive.
\begin{description}
\item[$\star$] {\bf $a^2\ge 0$}: The denominators in the integrands of
  \eqref{I} have zeros at $z\in \mathbb{Z}$. These are all compensated
  by the zeros of the numerators, except for the one at
  $z=0$. Therefore the integrands have a simple pole at
  $z=0$. Moreover, we have branch cuts as $z=-\delta\pm i\tau$,
  $\tau\in \left[|aR|,+\infty\right[$, which are induced by the branch
      cut of $\sqrt{z}$ at $z\in \mathbb{R}^-$. Thus, in this case we consider the contours of integration exhibited in figure \ref{fignegative} (left). Let us consider separately the several contributions:

\begin{figure}
\begin{picture}(0,0)(0,0)
\end{picture}
\centering\epsfig{file=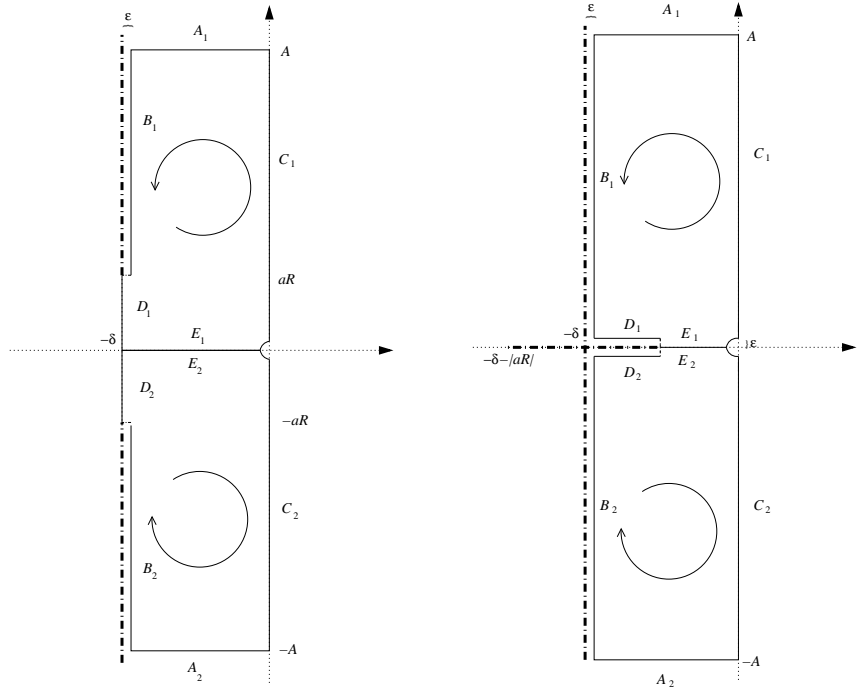,width=11.3cm}
\caption{Closed contours of integration: left/right figures correspond to the cases with $a^2\ge 0$ and $a^2<0$, respectively.}
\label{fignegative}
\end{figure}

\textit{Paths $A_1$ and $A_2$}: Integrate along $z=\pm iA+x$ respectively, with $x\in ]-\delta+\epsilon,0]$.  The exponential factor in the denominator will suppress these contributions when $A\rightarrow+\infty$. Thus
\[
\lim_{A\rightarrow +\infty}\left(I_{A_1}+I_{A_2}\right)=0 \ .
\]

\textit{Paths $B_1$ and $B_2$}: Take $A\rightarrow +\infty$ and integrate along $z=\pm i\tau-\delta+\epsilon$, with $\tau\in \left[aR,+\infty\right[$, respectively and $\epsilon$ infinitesimal; taking the limit $\epsilon\rightarrow 0$  one finds that the sum of these contributions is  non-vanishing only for $n$ odd,
\[
\lim_{\epsilon\rightarrow 0}\left(I_{B_1}+I_{B_2}\right)=\left\{ \barr{l}
\displaystyle{\int_{aR}^{+\infty} d\tau\dfrac{P_-(i\tau)\sqrt{\tau^2-a^2R^2}}{\exp(2\pi\tau)-1}} \ , \ \ \ \ n=2p+3 \ , \spa{0.4cm}\\ 0 \ , \ \ \ \ \  \ \ \ \ \ \ \ \ \ \ \ \ \ \ \ \ \  \ \ \ \ \ \ \ \ \ \ \ \ \ \ \ n=2p+4 \ . \earr \right. 
\]

\textit{Paths $D_1$ and $D_2$}:
Integrate along $z=\pm i\tau-\delta$, with $\tau\in [0,aR]$ respectively. One finds that the sum of these contributions in non-vanishing only for $n$ even,
\[
I_{D_1}+I_{D_2}=\left\{ \barr{l} 0 \ , \ \ \ \ \  \ \ \ \ \ \ \ \ \ \ \ \ \ \ \ \ \  \ \ \ \ \ \ \ \ \ \ \ \ \ \ \ 
 n=2p+3 \ , \spa{0.4cm}\\ \displaystyle{i\int_{0}^{aR} d\tau\dfrac{P_-(i\tau)\sqrt{a^2R^2-\tau^2}}{\exp(2\pi\tau)+1} \ ,  \ \ \ \ n=2p+4 \ .} \earr \right. 
\]

\textit{Paths $E_1$ and $E_2$}:
Integrate along $z=-t$, with \mbox{$t\in ]0,\delta[$,}
\[
I_{E_1}+I_{E_2}=-\int_{0}^{\delta}dy f_{\gamma}(y) \ . 
\]

\textit{Pole contribution at the origin}:
Integrate along $z=\epsilon e^{\pm i\theta}$ respectively, with $\theta\in [\pi,\pi/2]$ and $\epsilon$ infinitesimal, to obtain for the sum of these contributions is,
\[
\lim_{\epsilon\rightarrow 0}\left(I_{P_1}+I_{P_2}\right)=\dfrac{f_{0}(\delta)}{2} \ .
\]

Applying now Cauchy's theorem and substituting in \eqref{reg_abelplana} one obtains the manifestly real expression for the renormalised energy density (\ref{int1}).

\item[$\star$] {\bf $a^2< 0$}: The pole is the same as in the case $a^2\ge 0$, but the branch cuts are different. They are exhibited in  figure  \ref{fignegative} (right). Let us consider separately the several contributions:

\textit{Paths $A_1$ and $A_2$}: Same as in the case $a^2\ge 0$.

\textit{Paths $B_1$ and $B_2$}: Take $A\rightarrow +\infty$ and integrate along $z=\pm i\tau-\delta+\epsilon$, with $\tau\in \left[0,+\infty\right[$, respectively and $\epsilon$ infinitesimal; taking the limit $\epsilon\rightarrow 0$  one finds that the sum of these contributions is  non-vanishing only for $n$ odd,
\[
\lim_{\epsilon\rightarrow 0}\left(I_{B_1}+I_{B_2}\right)=\left\{ \barr{l}
\displaystyle{\int_{0}^{+\infty} d\tau\dfrac{P_-(i\tau)\sqrt{\tau^2-a^2R^2}}{\exp(2\pi\tau)-1}} \ , \ \ \ \ n=2p+3 \ , \spa{0.4cm}\\ 0 \ , \ \ \ \ \  \ \ \ \ \ \ \ \ \ \ \ \ \ \ \ \ \  \ \ \ \ \ \ \ \ \ \ \ \ \ \ \ \ \ n=2p+4 \ . \earr \right. 
\]

\textit{Paths $D_1$ and $D_2$}: Integrate along $z=\pm i\epsilon-\delta+t$, with $t\in [0,|a|R]$ respectively.  In the limit $\epsilon\rightarrow 0$ one finds that the sum of these contributions is
\[
\lim_{\epsilon\rightarrow 0}\left(I_{D_1}+I_{D_2}\right)=\left\{\begin{array}{ll}
\displaystyle{-\dfrac{1}{2}\int_{0}^{|a|R} dt P_-(t)\sqrt{-a^2R^2-t^2}\dfrac{\cos{\pi t}}{\sin{\pi t}} \ ,}& n=2p+3 \ ,\\
& \\
\displaystyle{\dfrac{1}{2}\int_{0}^{|a|R} dt P_-(t)\sqrt{-a^2R^2-t^2}\dfrac{\sin{\pi t}}{\cos{\pi t}}\ ,}& n=2p+4 \ .\end{array}
\right.
\]

\textit{Paths $E_1$ and $E_2$}:
Integrate along $z=-\delta+t$, with $t\in [|a|R,\delta[$. One finds a
    similar result as in the case $a^2\ge 0$,
\[
I_{E_1}+I_{E_2}=-\int_{|aR|}^{\delta}dy f_{\gamma}(y) \ . 
\]

\textit{Pole contribution at the origin}: This is the same as for $a^2\geq0$.

Applying now Cauchy's theorem and substituting in
\eqref{reg_abelplana} one obtains the manifestly real expression \eqref{int2}.

\end{description}

\end{document}